\documentclass[aps,pra,twocolumn,reprint,groupedaddress]{revtex4-2}

\usepackage{physics}
\usepackage{graphicx, color, graphpap}
\usepackage{amsmath}
\usepackage{amsfonts}
\usepackage[caption=false,subrefformat=parens,labelformat=parens]{subfig}
\usepackage{enumitem}
\usepackage{dsfont}

\DeclareMathOperator{\I}{I}
\DeclareMathOperator{\se}{H}

\DeclareMathOperator{\id}{id}
\usepackage[hyperindex]{hyperref}

\newenvironment{aeq}{\begin{equation}
\begin{aligned}
}{
\end{aligned}
\end{equation}}

\begin{document}

\title{Trusted detector noise analysis for discrete modulation schemes of \\ continuous-variable quantum key distribution}

\author{Jie Lin}
\affiliation{Institute for Quantum Computing and Department of Physics and Astronomy, University of Waterloo, Waterloo, Ontario, Canada N2L 3G1} 
\author{Norbert L\"utkenhaus}
\affiliation{Institute for Quantum Computing and Department of Physics and Astronomy, University of Waterloo, Waterloo, Ontario, Canada N2L 3G1}

\date{\today}

\begin{abstract}
Discrete-modulated continuous-variable quantum key distribution protocols are promising candidates for large-scale deployment due to the large technological overlap with deployed  modern optical communication devices. The security of discrete modulation schemes has previously analyzed in the ideal detector scenario in the asymptotic limit. In this work, we calculate asymptotic key rates against collective attacks in the trusted detector noise scenario. Our results show that we can thus cut out most of the effect of detector noise and obtain asymptotic key rates similar to those had we access to ideal detectors. 
\end{abstract}

\maketitle

\section{Introduction}
Quantum key distribution (QKD) \cite{Bennett1984, Ekert1991} is a key establishment protocol with the provable information-theoretic security. Various QKD protocols with different advantages have been proposed, analyzed and implemented. See e.g.  Refs. \cite{Scarani2009, Diamanti2015, Xu2020, Pirandola2019} for reviews. Continuous-variable (CV) QKD protocols \cite{Grosshans2002, Silberhorn2002,Grosshans2003a, Weedbrook2004} have competitive advantages in terms of massive deployment due to a significant overlap of devices used with those in the optical classical communications. Many experiments of CVQKD on both Gaussian modulation schemes such as Refs. \cite{Lodewyck2007, Jouguet2013, Qi2015, Soh2015, Huang2015, Huang2016, Zhang2020l} and discrete modulation schemes like Refs. \cite{Wittmann2010, Wang2013, Heim2014, Hirano2017, Laudenbach2019a} have been demonstrated. 

On one hand,  Gaussian modulation schemes are simpler to analyze theoretically than discrete modulation schemes,  and they give secret key rates close to the theoretical limits \cite{Takeoka2014,Pirandola2017}. On the other hand, continuous modulation itself is usually only approximated by a (relatively large) set of discrete modulation settings.  This approximation needs to be taken into account during the full security analysis (see e.g. \cite{Furrer2012, Jouguet2012, Kaur2019}). Moreover, as Gaussian modulation schemes often require more resources in terms of randomness and classical postprocessing resources, discrete modulation schemes thus offer further simplification of implementation. However, in previous experimental demonstrations of discrete modulation schemes, either only effective entanglement has been verified \cite{Wittmann2010, Heim2014}, which is a necessary precondition for QKD, or the security has been established only against a restricted subset of collective attacks  \cite{Wang2013, Hirano2017}. By now, there are asymptotic security proofs against arbitrary collective attacks for binary \cite{Zhao2009}, ternary \cite{Bradler2018} as well as quaternary modulation schemes and beyond \cite{Ghorai2019, Lin2019}. Previous proofs for a general discrete modulation scheme \cite{Ghorai2019, Lin2019} investigate the untrusted detector noise scenario where the imperfection of detectors can be controlled by Eve (and thus one can treat detectors as ideal). In reality, the amount of electronic noise of an off-the-shelf homodyne detector in a CVQKD experiment can be much higher than the channel excess noise. As a result, the key rate in the untrusted detector noise scenario drops very quickly to zero as the transmission distance increases. However, since detectors are securely located in Bob's laboratory where Eve is unable to access, it is reasonable to assume that Eve does not control detector imperfections especially those noise sources that are on the electronic circuitry, which is more remote from the quantum mechanical part of the signal detection. 

In this work, we extend our previous analysis \cite{Lin2019} to the trusted detector noise scenario where detector imperfections (detector inefficiency and electronic noise) are not accessible to Eve. We remark that Gaussian modulation schemes have been analyzed in the trusted detector noise scenario \cite{Lodewyck2007, Fossier2009, Usenko2016, Laudenbach2019b} and it is known that the effects of electronic noise and detector inefficiency on the key rates are not very significant in the trusted detector noise scenario compared to the ideal detector scenario under realistic experimental conditions. As we show in this work, this observation also holds for discrete modulation schemes. However, we emphasize that our analysis is not a trivial application of the method used for Gaussian modulation protocols and instead we adopt a different approach. The reason is that the previous method used in the Gaussian modulation protocols relies on the fact \cite{Navascues2006,Garcia-Patron2006} that Eve's optimal attacks for Gaussian modulation schemes correspond to Gaussian channels,  which make it easy to decouple the trusted detector noise from the channel noise when one looks at the covariance matrix. However, we cannot assume Gaussian channels here since Gaussian attacks are not expected to be optimal for discrete modulation schemes. In our analysis, based on a (commonly used) quantum optical model of the imperfect detector, we find its corresponding mathematical description in terms of positive operator-valued measure (POVM) and then use this POVM to construct observables corresponding to quantities that are measured experimentally. These observables are then used in our security proof. We also point out the crucial difference between our analysis and Ref. \cite{Namiki2018} for discrete modulation schemes: Our asymptotic analysis is valid against arbitrary collective attacks while Ref. \cite{Namiki2018} uses the Gaussian channel assumption and thus its security analysis \cite{Namiki2018} is restricted to Gaussian collective attacks.

Our main contributions of this work are finding a suitable POVM description of a noisy heterodyne detector and revising our previous analysis \cite{Lin2019} by using a new set of constraints from this POVM in the numerical key rate optimization problem \cite{Coles2016, Winick2018}. Similar to our previous analysis, this method is applicable to both direct reconciliation and reverse reconciliation schemes. Moreover, we study the postselection of data \cite{Silberhorn2002} in the trusted detector noise scenario. As a concrete example, we apply our method to the quadrature phase-shift keying scheme with heterodyne detection and focus on the reverse reconciliation scheme. Our analysis here is still restricted to the asymptotic regime against collective attacks and we make the same photon-number cutoff assumption as in the previous works \cite{Ghorai2019, Lin2019} to truncate the infinite-dimensional Hilbert space in order to perform the numerical calculation. From the numerical observation, we believe the results do not depend on the choice of cutoff when it is appropriately chosen. We direct the discussion about this assumption to Sec. III B of Ref. \cite{Lin2019} and leave it for future work to provide an analytical justification of this assumption beyond the numerical evidences. To extend our analysis to the finite-key regime, we remark that we have recently extended the numerical method of Ref. \cite{Winick2018} on which our analysis is based to include finite key analysis \cite{George2020}. However, there remain some technical challenges to solve before we can apply this method to this protocol and thus we leave the finite key analysis for future work.

The rest of paper is outlined as follows. In Sec. \ref{sec:background}, we review the protocol and proof method in Ref. \cite{Lin2019}. We then present a trusted detector noise model and the corresponding POVM description in Sec. \ref{sec:trustedNoise}. In Sec. \ref{sec:reformulation}, we modify the key rate optimization problem to take trusted detector noise into account. We discuss our simulation method in Sec. \ref{sec:simulation}. We show the simulation results without postselection in Sec. \ref{sec:results_nops} and with postselection in Sec. \ref{sec:results_ps}. Finally, we summarize the results and provide insights for future directions in Sec. \ref{sec:outlook}. We present technical details in the Appendices.

\section{Background}\label{sec:background}

Our key rate calculation in the trusted detector noise scenario uses a similar proof method as in our previous work \cite{Lin2019}; that is, we numerically perform the key rate optimization problem \cite{Winick2018} with a modified set of constraints. In particular, we discuss how to modify the key rate optimization problem in Sec. \ref{sec:reformulation} based on the POVM description of a noisy heterodyne detector in Sec. \ref{sec:trustedNoise}. To help understand this modification, we first review main ideas of the proof in Ref. \cite{Lin2019}.

For illustration, we focus on the quadrature phase-shift keying scheme with heterodyne detection. We remark that since the previous proof can be generalized to other discrete modulation schemes beyond four coherent states at the cost of more computational resources, our modified analysis for the trusted detector noise scenario can also be generalized in the same way. Moreover, one can apply a similar idea presented in this paper to study the homodyne detection scheme in the presence of trusted detector noise.

\subsection{Quadrature phase-shift keying protocol}
To begin with, we review the quadrature phase-shift keying (QPSK) scheme with heterodyne detection. The quantum part of the protocol consists of many repetitions of the following two steps:
 (1) Alice obtains a uniform random number $x \in \{0,1,2,3\}$, selects the state $\ket{\alpha_x} = \ket{\alpha e^{\frac{ix\pi}{2}}}$ from the set $\{\ket{\alpha}, \ket{i\alpha},\ket{-\alpha}, \ket{-i\alpha}\}$ according to the value of $x$, and sends it to Bob. (2) Bob applies the heterodyne detection to the received state and obtains a measurement outcome $y \in \mathbb{C}$. 
 
After the quantum communication phase of the protocol, they proceed with the classical postprocessing part of the protocol including announcement, sifting, parameter estimation, key map (with discretization), error correction and privacy amplification. In particular, the parameter estimation step is done according to the key rate optimization problem in Eq. (\ref{eq:optimization_reformulated}) discussed later. As the classical part is similar to other CVQKD protocols and is not the focus of our discussion, we highlight only the key map step below for our discussion and skip the details of the remaining classical postprocessing procedures here. We direct readers to Ref. \cite{Lin2019} for a more detailed description.
 
In the case of reverse reconciliation, for each measurement outcome $y$ written as $y = \abs{y} e^{i \theta}$, where $\theta \in [-\frac{\pi}{4},\frac{7\pi}{4})$, Bob obtains a discretized value $z$ according to the following rule:
\begin{equation}
z =\begin{cases}
j, & \text{if }\theta \in \Big[\frac{(2j-1)\pi}{4}, \frac{(2j+1)\pi}{4}\Big)\ \ \text{and} \ \abs{y} \geq \Delta_a\\
\perp, & \text{otherwise},
\end{cases}
\end{equation}where $j \in\{0,1,2,3\}$ and $\Delta_a$ is a postselection parameter that needs to be optimized for the selected protocol and experimental parameters \footnote{In our previous work \cite{Lin2019}, we also considered a postselection parameter $\Delta_p$ related to the phase of the measurement outcome. However, when we performed simulations with this postselection parameter, we did not obtain any noticeable advantage. Thus, we omit the introduction of this parameter in this work.}. A protocol without postselection corresponds to setting $\Delta_a= 0$. 

To perform the postselection of data in combination of reverse reconciliation, Bob announces positions where he obtains the value $\perp$. After removing the positions related to the value $\perp$, Alice's string $\vec{\mathbf{X}}$ consists of her random number $x$'s in the remaining positions, and Bob's raw key string $\vec{\mathbf{Z}}$ consists of his discretized outcome $z$'s left. (Alternatively, they may choose to announce and keep positions related to the value $\perp$ and let the privacy amplification subprotocol effectively remove those positions.) Alice and Bob may decide to recast their strings to binary strings before or during the error correction step depending on their choice of the error-correction code. For the consistency of our presentation, we use the alphabet $\{0,1,2,3\}$ and let $\mathbf{X}$ and $\mathbf{Z}$ denote the single-round version of $\vec{\mathbf{X}}$ and $\vec{\mathbf{Z}}$, respectively.

\begin{figure}[h]
\includegraphics[width=0.9\linewidth]{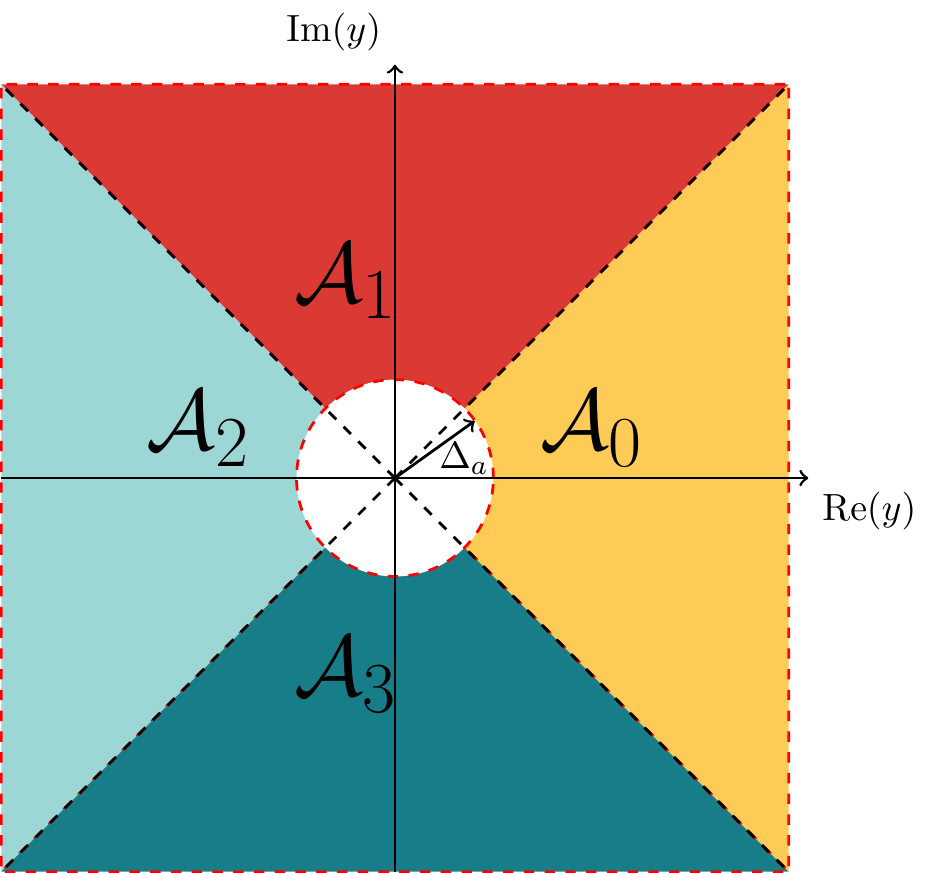}
\caption{\label{fig:keymap}Key map for quadrature phase-shift keying scheme in terms of Bob's measurement outcome $y \in \mathbb{C}$. Each colored region $\mathcal{A}_j$ corresponds to a discretized key value $j$. The measurement outcome in the central disk with a radius $\Delta_a$ is discarded during the postselection of data and is mapped to the symbol $\perp$. }
\end{figure}

\subsection{Review of security proof method}\label{sec:review}

\subsubsection{Source-replacement scheme}
The first step of our security proof is to apply the source-replacement scheme \cite{Bennett1992, Grosshans2003b, Curty2004, Ferenczi2012} to obtain an equivalent entanglement-based scheme for the given prepare-and-measure protocol. Then we proceed to prove the security of the entanglement-based scheme. 

Given Alice's state ensemble $\{\ket{\alpha_x}, p_x\}$ (where $p_x = \frac{1}{4}$ for this protocol) for her preparation in the prepare-and-measure scheme, Alice effectively prepares a bipartite state $\ket{\Psi}_{AA'}$ in the source-replacement scheme, which is defined as
\begin{equation}
\ket{\Psi}_{AA'} = \sum_{x=0}^3 \sqrt{p_x} \ket{x}_{A} \ket{\alpha_x }_{A'}, 
\end{equation}
where $\{\ket{x}\}$ is an orthonormal basis for register $A$. Then Alice sends the register $A'$ to Bob via an insecure quantum channel and keeps register $A$ for her measurement described by the POVM $M^A=\{M^A_x = \dyad{x}{x}:x \in \{0,1,2,3\}\}$. The quantum channel that maps register $A'$ to Bob's register $B$ is described by a completely positive (CP) trace-preserving (TP) map, $\mathcal{E}_{A'\rightarrow B}$ and is assumed to be under Eve's control. Thus, Alice and Bob's joint state $\rho_{AB}$ before their measurements is 
\begin{aeq}
\rho_{AB} = (\id_A  \otimes \mathcal{E}_{A'\rightarrow B}) (\dyad{\Psi}{\Psi}_{AA'}),
\end{aeq}where $\id_A$ is the identity channel on Alice's system $A$.

When Alice performs a local measurement using her POVM $\{M^A_x\}$ on register $A$ and obtains an outcome $x$, she effectively sends the coherent state $\ket{\alpha_x}$ to Bob. Bob's received state $\rho_B^{x}$ conditioned on Alice's choice of $x$ is
\begin{equation}\label{eq:conditional_state_B}
\rho_B^{x} = \frac{1}{p_x} \Tr_A[\rho_{AB} (\dyad{x}{x}_A \otimes \mathds{1}_B)].
\end{equation}
Bob applies his POVM $M^B=\{M^B_y\}$ to register $B$ to obtain his measurement outcomes. In the case of untrusted detector noise (or ideal heterodyne detector), the POVM of the heterodyne detection is $\{E_{y} =\frac{1}{\pi} \dyad{y}{y}: y \in \mathbb{C}\},$ where $\ket{y}$ denotes a coherent state with complex amplitude $y$.

\subsubsection{Key rate optimization}
The next step is to formulate the key rate optimization problem for the entanglement-based scheme. One can rewrite the well-known Devetak-Winter formula \cite{Devetak2005} into the following form \cite{Coles2016, Winick2018}
\begin{aeq}\label{eq:keyrate_objective}
R^{\infty} = \min_{\rho_{AB} \in \mathbf{S}}D\Big(\mathcal{G}(\rho_{AB}) || \mathcal{Z}[\mathcal{G}(\rho_{AB})]\Big) - p_{\text{pass}} \delta_{EC},
\end{aeq}where $\delta_{EC}$ is the actual amount of information leakage per signal pulse in the error-correction step, $D(\rho||\sigma) = \Tr(\rho \log_2 \rho) - \Tr(\rho \log_2 \sigma)$ is the quantum relative entropy between two (subnormalized) density operators $\rho$ and $\sigma$, $\mathcal{G}$ is a CP, trace nonincreasing map for postprocessing and $\mathcal{Z}$ is a pinching quantum channel for accessing results of the key map. The set $\mathbf{S}$ contains all density operators compatible with experimental observations. A more detailed discussion about the map $\mathcal{G}$ can be found in Appendix A of Ref. \cite{Lin2019}. For the reverse reconciliation scheme, we can express the cost of error correction $\delta_{EC}$ by
\begin{aeq}\label{eq:delta_ec_RR}
\delta_{\text{EC}} &=   \se(\mathbf{Z}) - \beta \I(\mathbf{X};\mathbf{Z}),
\end{aeq}where $\se(\mathbf{Z})$ is the Shannon entropy of the raw key $\mathbf{Z}$, $\beta$ is the reconciliation efficiency of the chosen error-correction code, and $\I(\mathbf{X};\mathbf{Z})$ is the classical mutual information between $\mathbf{X}$ and $\mathbf{Z}$.

Before we review the set of constraints as well as $\mathcal{G}$ and $\mathcal{Z}$ maps  for the quadrature phase-shift keying scheme, we start with basic definitions. Given the annihilation operator $\hat{a}$ and creation operator $\hat{a}^{\dagger}$ of a single-mode state with the usual commutation relation $[\hat{a}, \hat{a}^{\dagger}]=\mathds{1}$, we define the quadrature operators $\hat{q}$ and $\hat{p}$, respectively, as
\begin{aeq}\label{eq:quadrature_ops}
\hat{q} &= \frac{1}{\sqrt{2}} (\hat{a}^{\dagger} + \hat{a}), \; \; \;
\hat{p} = \frac{i}{\sqrt{2}} (\hat{a}^{\dagger} - \hat{a}).
\end{aeq}They obey the commutation relation $[\hat{q},\hat{p}] = i \mathds{1}$. 
 To utilize the second-moment observations $\langle \hat{q}^2\rangle$ and $\langle \hat{p}^2\rangle$ to constrain $\rho_{AB}$, we previously defined the following two operators $\hat{n} = \frac{1}{2}( \hat{q}^2 + \hat{p}^2 - \mathds{1})=\hat{a}^{\dagger}\hat{a}$ and $\hat{d} = \hat{q}^2 - \hat{p}^2 = \hat{a}^2 +  (\hat{a}^{\dagger})^2$ \cite{Lin2019}. The relation between these observables and the heterodyne detection POVM is highlighted in Sec. \ref{sec:reformulation_untrusted}.

For the untrusted detector noise (or ideal heterodyne detector) scenario, the key rate optimization problem \cite{Lin2019} is
\begin{aeq}\label{eq:optimization}
\text{minimize }\; & D\big(\mathcal{G}(\rho_{AB}) || \mathcal{Z}[\mathcal{G}(\rho_{AB})]\big)\\
\text{subject to }\; & \\
& \Tr[\rho_{AB} (\dyad{x}{x}_A \otimes \hat{q})] = p_x \langle \hat{q} \rangle_x, \\
& \Tr[\rho_{AB} (\dyad{x}{x}_A \otimes \hat{p})] = p_x \langle \hat{p} \rangle_x,\\
& \Tr[\rho_{AB} (\dyad{x}{x}_A \otimes \hat{n})] = p_x \langle \hat{n} \rangle_x, \\
& \Tr[\rho_{AB} (\dyad{x}{x}_A \otimes \hat{d})] = p_x \langle \hat{d} \rangle_x, \\
& \Tr[\rho_{AB}] = 1,\\
& \Tr_B [\rho_{AB}] = \sum_{i,j=0}^3 \sqrt{p_i p_j} \bra{\alpha_j}\ket{\alpha_i} \dyad{i}{j}_{A}, \\ 
& \rho_{AB} \geq 0, 
\end{aeq}where the index $x$ runs over the set $\{0,1,2,3\}$ and $\langle \hat{q} \rangle_x, \langle \hat{p} \rangle_x, \langle \hat{n} \rangle_x$, and $ \langle\hat{d} \rangle_x$ denote the corresponding expectation values of operators $\hat{q}, \hat{p}, \hat{n}$, and $\hat{d}$ for the conditional state $\rho_B^x$, respectively.

As indicated in Fig. \ref{fig:keymap}, the protocol can perform postselection of data. To perform postselection, we defined the region operators in Ref. \cite{Lin2019} as
\begin{aeq}\label{eq:region_operator}
R_{j} &= \frac{1}{\pi}\int_{\Delta_a}^{\infty} \int_{\frac{(2j-1)\pi}{4}}^{\frac{(2j+1)\pi}{4}}  r \dyad{r e^{i\theta}}{r e^{i\theta}} \; d\theta \; dr
\end{aeq}for $j \in \{0,1,2,3\}$. The area of integration for each operator corresponds to a region shown in Fig. \ref{fig:keymap}.

The postprocessing map $\mathcal{G}$ in the reverse reconciliation scheme is given by $\mathcal{G} (\sigma) = K\sigma K^{\dagger}$  for any input state $\sigma$, where the Kraus operator $K$ is
\begin{aeq}\label{eq:kraus_op}
K = \sum_{z=0}^3  \ket{z}_{R} \otimes \mathds{1}_A \otimes (\sqrt{R_{z}})_B,
\end{aeq}where $\{\ket{0}_R, \ket{1}_R, \ket{2}_R, \ket{3}_R \}$ is the standard basis for register $R$. The pinching quantum channel $\mathcal{Z}$ is given by projections $\{\dyad{j}{j}_{R} \otimes \mathds{1}_{AB}: j \in \{0, 1,2,3\}\}$ as
\begin{aeq}
\mathcal{Z}(\sigma) = \sum_{j =0}^{3} (\dyad{j}{j}_{R} \otimes \mathds{1}_{AB}) \sigma (\dyad{j}{j}_{R} \otimes \mathds{1}_{AB} ).
\end{aeq}

\section{Noisy heterodyne detection}\label{sec:trustedNoise}
In this section, we present one physical model for a noisy heterodyne detector and give the corresponding POVM description. We start with a slightly more general model and then we make a simplification for the ease of calculation at the end of this section. This simplified model then reduces to a model commonly used in the literature.  
\subsection{Trusted detector noise model}
\begin{figure}[h]
\includegraphics[width=\linewidth]{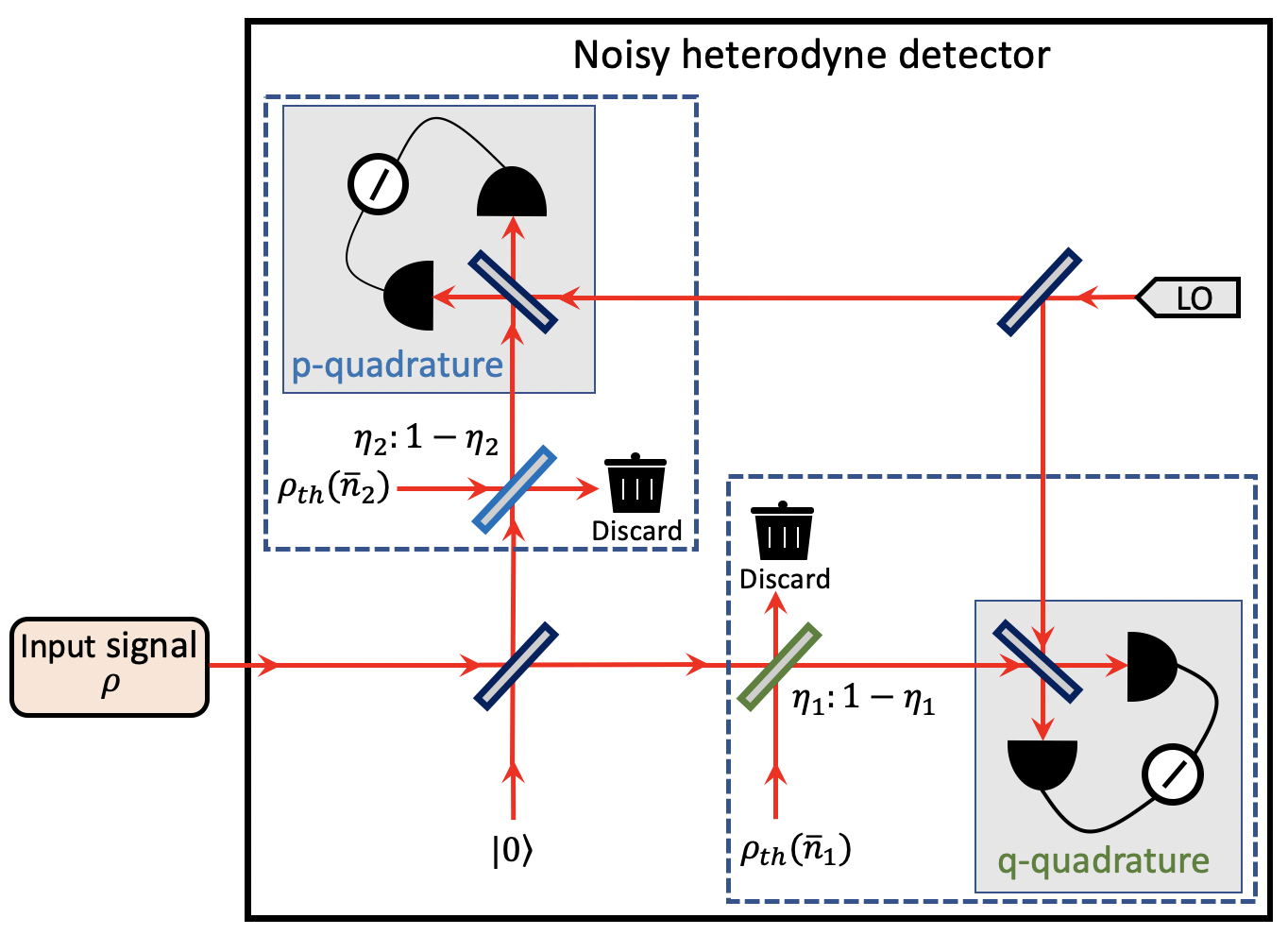}
\caption{\label{fig:noisy_het}Physical model for a noisy heterodyne detector. The homodyne detector for the $q$ quadrature measurement has detector efficiency $\eta_1$ and electronic noise $\nu_1$. The homodyne detector for the $p$ quadrature measurement has detector efficiency $\eta_2$ and electronic noise $\nu_2$. The notation $\rho_{th}(\bar{n})$ stands for a thermal state with a mean photon number $\bar{n}$. In particular, $\bar{n}_1 = \frac{\nu_{1}}{2(1-\eta_1)}$ and $\bar{n}_2 = \frac{\nu_{2}}{2(1-\eta_2)}$ (see main text for more explanations). beam-splitters are 50:50 unless specified otherwise. Each homodyne detector inside a gray box is ideal. Each dashed box encloses the physical model for a noisy homodyne detector. LO stands for local oscillator.  }
\end{figure}

As a heterodyne detector consists of two homodyne detectors and a beam-splitter, we consider imperfections in each homodyne detector. A homodyne detector may have nonunity detector efficiency and also have some amount of electronic noise which is the additional noise introduced to the measured data by its electronic components. In an experiment, one is able to measure the amount of electronic noise and the value of detector efficiency by a calibration routine. To model a realistic homodyne detector with nonunity detector efficiency and some amount of electronic noise, we use a quantum optical model which is used in Refs. \cite{Lodewyck2007, Fossier2009, Usenko2016, Namiki2018, Laudenbach2019b}, although the source of this electronic noise is in the actual electronics part of the detector. An alternative view of the electronic noise is that we can think about the detector as being a perfect detector followed by some classical postprocessing of the data, which adds noise. One should note that in a trusted device scenario, the characterization of the actual noise should be experimentally verified. Our physical model is chosen for convenience of calculating the POVM of the actual measurement. We depict this physical model of a noisy heterodyne detector in Fig. \ref{fig:noisy_het}. In this diagram, we consider a more general case where two homodyne detectors have different imperfections. We label the efficiency of the homodyne detector used for $q$ quadrature measurement as $\eta_1$ and its electronic noise as $\nu_1$ (expressed in shot noise units). Similarly, the efficiency of the homodyne detector used for $p$ quadrature measurement is labeled as $\eta_2$ and its electronic noise is labeled as $\nu_2$. 

Since our treatment for each homodyne detector in this heterodyne setup is the same, we take one homodyne detector (shown in each dashed box in Fig.  \ref{fig:noisy_het}) as an example and treat the other one similarly by using its corresponding efficiency and electronic noise. An imperfect homodyne detector with its efficiency $\eta_j < 1$ and electronic noise $\nu_j \geq 0$ (for $j=1$ or $2$) can be modeled by a beam-splitter placed before a perfect homodyne detector with the following specification. (1) The ratio of transmission to reflection of this beam-splitter is $\eta_j: 1-\eta_j$. (2) One input port of this beam-splitter is the signal pulse and the other input port is a thermal state used to model electronic noise, which is equivalent to sending one mode of a two-mode squeezed vacuum state (EPR state) to the beam-splitter. Each quadrature's variance of this ancillary thermal state is related to the value of electronic noise $\nu_j$. More specifically, it is $[1+\nu_j/(1-\eta_j)]N_0$ \cite{Lodewyck2007}, where $N_0=1/2$ denotes the shot-noise variance. In Fig. \ref{fig:noisy_het}, we choose to parametrize the thermal state in terms of its mean photon number as $\bar{n}_j=\frac{\nu_j}{2(1-\eta_j)}$ instead of the variance of each quadrature, which is convenient for writing of expressions in later sections \footnote{The electronic noise $\nu_j$ is the thermal noise added by the detection electronics. In the quantum mechanical model of the detector shown in each dashed box of Fig. \ref{fig:noisy_het} , the electronic noise is modeled by an ancillary thermal state added to the second input port of the beam-splitter that models the detector efficiency. Since the value of electronic noise is unaffected by the detector efficiency, to simulate the desired amount of noise before this beam-splitter, one then needs to scale it by the reflectance of the beam-splitter which is $1-\eta_j$. As the variance of a thermal state with a mean photon number $\bar{n}$ is $(1+2\bar{n})N_0$, one can easily see that the mean photon number of this ancillary thermal state is $\bar{n}_j=\frac{\nu_j}{2(1-\eta_j)}$. }. We note that this way of modeling electronic noise is valid when $\eta_j \neq 1$. Furthermore, we assume $\eta_j \neq 0$. That is, we consider the case $\eta_j \in (0,1)$, which is the case of a realistic detector of our interest.

In the next section, we derive the POVM corresponding to this detector model. We then choose to consider a simplified scenario where these two homodyne detectors are identical for the purpose of illustration and the ease of numerical calculation. That is, we later assume they both have the same detector efficiency $\eta_1=\eta_2=:\eta_d$ and the same electronic noise $\nu_1=\nu_2=:\nu_{\text{el}}$.

\subsection{POVM description}
We use the Wigner function formulation to find the POVM $\{G_y: y \in \mathbb{C}\}$ corresponding to this noisy heterodyne detector model. When two homodyne detectors give two real numbers $q_s$ and $p_s$ for $q$ and $p$ quadrature measurements, we label the outcome as $y = q_s + i p_s$. By considering $\Tr(\rho G_y)$ for an arbitrary input density operator $\rho$ to the noisy heterodyne detector, we are able to find the Wigner function $W_{G_y}$ of the POVM element $G_y$ as
\begin{aeq}\label{eq:wigner_Gy_main}
W_{G_y}(\gamma) =&  \frac{1}{\sqrt{\eta_1\eta_2}\pi}\frac{2}{\pi}\frac{1}{\sqrt{1+\frac{2(1-\eta_1+\nu_{1})}{\eta_1}}}\frac{1}{\sqrt{1+\frac{2(1-\eta_2+\nu_{2})}{\eta_2}}}\\
 &\times \exp(\frac{-2  [\Re(\gamma)-\frac{1}{\sqrt{\eta_1}}\Re(y)]^2}{1+\frac{2(1-\eta_1+\nu_{1})}{\eta_1}})\\
 & \times \exp(\frac{-2 [\Im(\gamma)- \frac{1}{\sqrt{\eta_2}}\Im(y)]^2}{1+\frac{2(1-\eta_2+\nu_{2})}{\eta_2}}).
\end{aeq}

By comparing this Wigner function with that of a displaced squeezed thermal state, we can identify that the POVM element $G_y$ is a projection onto a displaced squeezed thermal state up to a prefactor $\frac{1}{\sqrt{\eta_1\eta_2}\pi}$. We give a full derivation of this Wigner function and the explicit parameters for displacement, squeezing and thermal state mean photon number in terms of detector parameters $\eta_1, \eta_2, \nu_1$ and $\nu_2$ in Appendix \ref{app:povm}. 

For the rest of the paper, we restrict our discussion to a simpler scenario where we assume both homodyne detectors have the same imperfection for the ease of numerical calculation and for the purpose of illustration. We discuss how to perform the calculation in the general case in Appendix \ref{app:representation}. In this simple case, we set $\eta_1 =\eta_2 =\eta_d$ and $\nu_{1} = \nu_{2} = \nu_{\text{el}}$ in Eq. (\ref{eq:wigner_Gy_main}).  This equation is simplified to be
\begin{aeq}
W_{G_y}(\gamma) &=  \frac{1}{\eta_d\pi}\frac{2}{\pi}\frac{1}{1+\frac{2(1-\eta_d + \nu_{\text{el}})}{\eta_d}}  \exp(\frac{-2 \abs{\gamma- \frac{y}{\sqrt{\eta_d}}}^2}{1+\frac{2(1-\eta_d + \nu_{\text{el}})}{\eta_d}}).
\end{aeq}One can observe that it is the Wigner function of a displaced thermal state apart from the prefactor $ 1/(\eta_d\pi)$. Therefore, the POVM element $G_y$ in this case is a scaled projection onto a displaced thermal state. More precisely,
\begin{aeq}\label{eq:noisy_het_povm}
G_{y} = \frac{1}{\eta_d\pi} \hat{D}(\frac{y}{\sqrt{\eta_d}})\rho_{\text{th}}(\frac{1-\eta_d + \nu_{\text{el}}}{\eta_d})\hat{D}^{\dagger}(\frac{y}{\sqrt{\eta_d}}),
\end{aeq}where $\hat{D}(\frac{y}{\sqrt{\eta_d}})$ is the displacement operator with the amount of displacement $y/\sqrt{\eta_d}$ and $\rho_{\text{th}}(\frac{1-\eta_d + \nu_{\text{el}}}{\eta_d})$ is a thermal state with the mean photon number $(1-\eta_d + \nu_{\text{el}})/\eta_d$, which can be expressed in the photon-number basis as
\begin{aeq}
\rho_{\text{th}}(\bar{n}) = \sum_{n=0}^{\infty} \frac{\bar{n}^n}{(1+\bar{n})^{n+1}}\dyad{n}{n}.
\end{aeq}Later in Sec. \ref{sec:reformulation}, we need to express operators defined in terms of POVM elements $G_y$'s in the photon-number basis for the numerical key rate calculation. Analytical expressions of matrix elements $\bra{m}G_y\ket{n}$ are known in the literature \cite{Mollow1967} and shown in Appendix \ref{app:representation}.

Let us end this section with a few remarks about the simplification considered here. Firstly, as we later define operators involving integrals of POVM elements $G_y$'s and need to find their matrix representations in the photon-number basis for the numerical key rate calculation, we are able to find efficiently computable analytical expressions for these operators under this simplification. Without this simplification, one may need to perform some numerical integrations. We emphasize that the principles presented in this work also hold for the general case and we choose to present results based on this simplified case for the ease of calculation. Secondly, with this simplification, our detector model is then optically equivalent to the detector model used in other works \cite{Fossier2009, Laudenbach2019b}. Thirdly, if two homodyne detectors in the heterodyne detection scheme do not have the same imperfection, one can instead use the POVM in the general case by following the procedure outlined in Appendix \ref{app_sec:general_case} despite being more numerically challenging.

\section{Key rate optimization problem}\label{sec:reformulation}
We start with a reformulation of the optimization problem in Eq. (\ref{eq:optimization}) in the untrusted detector noise scenario which serves as a basis for our modification in the trusted detector noise scenario. The purpose of this reformulation is that once we substitute the POVM of the noisy heterodyne detector in place of the one for the ideal heterodyne detector, we can easily formulate the optimization problem in the trusted detector noise scenario. Specifically, we change Bob's POVM $\{M^B_y\}$ from the ideal heterodyne detection $\{E_{y}=\frac{1}{\pi} \dyad{y}{y}\}$ to the POVM description of the noisy heterodyne detection $\{G_y\}$ found in Eq. (\ref{eq:noisy_het_povm}). Moreover, compared with our previous work \cite{Lin2019}, some constraints are modified to match with how data are processed in a typical experiment. 
\subsection{Reformulation of the optimization problem in the untrusted detector noise scenario}\label{sec:reformulation_untrusted}
We reconsider the key rate optimization problem in the untrusted detector noise scenario by rewriting region operators in Eq. (\ref{eq:region_operator}) and observables in Eq. (\ref{eq:optimization}) in terms of the POVM of an ideal heterodyne detector $\{E_y\}$. In the case of ideal heterodyne detection, the POVM description of Bob's measurement $\{M^B_y\}$ is $M^B_y=E_y =\frac{1}{\pi}\dyad{y}{y}$, the projection onto a coherent state $\ket{y}$. By writing $y=re^{i\theta}$ in the polar coordinate and integrating over the corresponding region $\mathcal{A}_j$, we obtain Eq. (\ref{eq:region_operator}). If we rewrite Eq. (\ref{eq:region_operator}) in terms of $M^B_y$, we see region operators $R_j$'s are defined by 
\begin{aeq}\label{eq:region_operator_general}
R_j = \int_{y \in \mathcal{A}_j} M^B_{y}d^2 y,
\end{aeq}where the region of integration $\mathcal{A}_j$ in the complex plane is shown in Fig. \ref{fig:keymap} and $d^2 y = d \Re(y) d \Im(y)$. 

From the heterodyne detection, we obtain a probability density function $P(y)$ for the outcome $y \in \mathbb{C}$. (We obtain such a probability density function for each conditional state $\rho_B^x$. While it is more proper to denote this conditional probability density function as $P(y|x)$, for simplicity of notation in this section, we use $P(y)$.) When the heterodyne detector is ideal, this probability density function is the Husimi $Q$ function. In particular, as discussed in our previous work \cite{Lin2019}, the expectation values of operators $\hat{q}, \hat{p}, \hat{n}$ and $\hat{d}$ defined in Sec. \ref{sec:review} are related to the $Q$ function via 
\begin{aeq}\label{eq:previous_observables}
\langle \hat{q} \rangle_x  &= \frac{1}{\sqrt{2}}\int (y+ y^*) Q_x(y)  d^2 y,\\
\langle \hat{p} \rangle_x  &= \frac{i}{\sqrt{2}}\int (y^* - y) Q_x(y)  d^2 y,\\
\langle \hat{n} \rangle_x  &=\int ( \abs{y}^2-1) Q_x(y)  d^2 y,\\
\langle \hat{d} \rangle_x  &=\int [y^2 + (y^*)^2] Q_x(y)  d^2 y,
\end{aeq}where the subscript $x$ labels the conditional state $\rho_B^x$.

In general, one may be interested in a quantity like $\int f(y, y^*)P(y) d^2 y$ where $f(y, y^*)$ is a real-valued function on $y$ and $y^*$ such that the integral converges. Such a quantity can be described as the expectation value of an observable that is defined in the following way 
\begin{equation}\label{eq:observables_general}
\hat{O} = \int f(y, y^*) M^B_{y} d^2 y
\end{equation}since
\begin{aeq}\label{eq:general_expectation}
\Tr[\rho \; \hat{O}] 
&= \int d^2 y \; f (y, y^*)  \Tr(\rho M^B_{y})\\
&= \int  d^2 y \; f(y, y^*)P(y).
\end{aeq}In other words, operators constructed in this way correspond to expectation values $\int f(y, y^*)P(y) d^2 y$ obtained in an experiment. By comparing Eq. (\ref{eq:general_expectation}) to Eq. (\ref{eq:previous_observables}) and identifying $P(y)$ by $Q_x(y)$, we observe the following choices of $f(y,y^*)$ for $\hat{q}$, $\hat{p}$, $\hat{n}$ and $\hat{d}$:
\begin{aeq}
  \hat{q} \longleftrightarrow & \; f(y, y^*) = \frac{y + y^*}{\sqrt{2}}, \\
  \hat{p} \longleftrightarrow & \; f(y, y^*) = \frac{i(y^* - y)}{\sqrt{2}},\\
   \hat{n} \longleftrightarrow & \; f(y, y^*) = \abs{y}^2 - 1,\\
   \hat{d} \longleftrightarrow  & \; f(y, y^*) = y^2 + (y^*)^2.
\end{aeq}We remark that this way of defining these observables corresponds to the antinormally ordered expansion of operators \cite{Cahill1969a, Cahill1969b}.

\subsection{Revised optimization problem in the trusted detector noise scenario}
In Ref. \cite{Lin2019}, we chose observables $\{\hat{O}\}= \{\hat{q}, \hat{p}, \hat{n}, \hat{d}\}$ by using $M^B_y = E_y$ in Eq. (\ref{eq:observables_general}) for the untrusted detector noise scenario. In this work, we change to a new set of observables $ \{\hat{q}, \hat{p}, \hat{n} + \hat{d}/2 +\mathds{1} , \hat{n} - \hat{d}/2 +\mathds{1} \}$, which gives the same key rates as the old one since the last two observables in this new set are linear combinations of observables $\hat{n}$ and $\hat{d}$ as well as the identity operator. This new set of observables corresponds to the set of $\{f(y,y^*)\} = \{\sqrt{2}\Re(y), \sqrt{2}\Im(y), 2 \Re(y)^2, 2\Im(y)^2\}$ \footnote{Due to our definition of quadrature operators, we include the factor $\sqrt{2}$ so that we can simply enter values reported in an experiment using shot noise units as expectation values of corresponding observables.}. The sole purpose of this change compared with Ref. \cite{Lin2019} is to make the data postprocessing in an agreement with the typical classical postprocessing in an experiment. That is, in an experiment, when a heterodyne detection gives two real numbers $q_s$ and $p_s$ which we set $\Re(y) = q_s$ and $\Im(y)=p_s$, one usually computes variances of $\Re(y)$ and $\Im(y)$ by computing the expectation values of $\Re(y)^2$ and $\Im(y)^2$ in addition to expectation values of $\Re(y)$ and $\Im(y)$.

In the trusted detector noise scenario, we need to substitute $M^B_y$ in Eqs. (\ref{eq:region_operator_general}) and (\ref{eq:observables_general}) by $G_y$. To distinguish operators defined in this way from the first and second moment of quadrature operators $\hat{q}$ and $\hat{p}$, we call first-moment observables $\hat{F}_Q$ and $\hat{F}_P$ and second-moment observables $\hat{S}_Q$ and $\hat{S}_P$. More explicitly, they are defined as
\begin{aeq}\label{eq:observables}
\hat{F}_Q &= \int \frac{y + y^*}{\sqrt{2}} G_y d^2y,\\
\hat{F}_P &= \int \frac{i(y^* - y)}{\sqrt{2}}G_y d^2y,\\
\hat{S}_Q &= \int  (\frac{y+ y^*}{\sqrt{2}})^2  G_y d^2y,\\
\hat{S}_P &= \int  [\frac{i(y^* - y)}{\sqrt{2}}]^2 G_y d^2y.
\end{aeq} 

Then the revised key rate optimization problem becomes
\begin{aeq}\label{eq:optimization_reformulated}
\text{minimize }\; & D\big(\mathcal{G}(\rho_{AB}) || \mathcal{Z}[\mathcal{G}(\rho_{AB})]\big)\\
\text{subject to }\; & \\
& \Tr[\rho_{AB} (\dyad{x}{x}_A \otimes \hat{F}_Q)] = p_x \langle \hat{F}_Q \rangle_x, \\
& \Tr[\rho_{AB} (\dyad{x}{x}_A \otimes \hat{F}_P)] = p_x \langle \hat{F}_P \rangle_x,\\
& \Tr[\rho_{AB} (\dyad{x}{x}_A \otimes \hat{S}_Q)] = p_x \langle \hat{S}_Q \rangle_x, \\
& \Tr[\rho_{AB} (\dyad{x}{x}_A \otimes \hat{S}_P)] = p_x \langle \hat{S}_P \rangle_x, \\
& \Tr[\rho_{AB}] = 1,\\
& \Tr_B [\rho_{AB}] = \sum_{i,j=0}^3 \sqrt{p_i p_j} \bra{\alpha_j}\ket{\alpha_i} \dyad{i}{j}_{A}, \\ 
& \rho_{AB} \geq 0, 
\end{aeq}where the index $x$ runs over the set $\{0,1,2,3\}$ and the Kraus operator for the postprocessing map $\mathcal{G}$ has the same form as in Eq. (\ref{eq:kraus_op}) but now with the region operators defined in terms of $G_y$'s in Eq. (\ref{eq:region_operator_general}). 

In Appendix \ref{app:representation}, we discuss how to represent these operators in the photon-number basis. Combining with the photon-number cutoff assumption (i.e. $\rho_{AB} = (\mathds{1}_A \otimes \Pi_{N}) \rho_{AB} (\mathds{1}_A \otimes \Pi_{N})$, where $N$ is the cutoff photon number and $\Pi_{N}$ is the projection onto the subspace spanned by the photon-number states from $0$ to $N$ photons), we can directly solve this key rate optimization problem in Eq. (\ref{eq:optimization_reformulated}) numerically. We direct readers to Sec. IV B of Ref. \cite{Lin2019} for the discussion about the numerical algorithm for the optimization problem and its performance.

\section{Simulation method}\label{sec:simulation}
In an experiment, the expectation values shown in the optimization problem in Eq. (\ref{eq:optimization_reformulated}) can be obtained from some suitable postprocessing of noisy heterodyne detection results. Without doing experiments, we perform simulations of a corresponding experiment with a noisy heterodyne detector to obtain those expectation values. With these values specified, one can solve the key rate optimization problem using a numerical convex optimization package to obtain numerical results. We emphasize that our security proof technique does not depend on the specific channel model used for the simulation.

\subsection{Channel model for simulation}
To understand how the protocol behaves in the trusted detector noise scenario, we simulate the quantum channel by using a realistic physical channel in an honest implementation of the protocol.  A realistic physical channel in the context of the optical fiber communication can be modeled by a phase-invariant Gaussian channel with the transmittance $\eta_t$ and excess noise $\xi$. In a typical fiber for optical communication, the attenuation coefficient is 0.2 dB/km and thus $\eta_t = 10^{-0.02 L}$ for a distance $L$ in kilometers. The excess noise $\xi$  is defined as
\begin{aeq}
\xi = \frac{(\Delta q_{\text{obs}})^2}{(\Delta q_{\text{vac}})^2} -1,
\end{aeq}where $(\Delta q_{\text{vac}})^2=N_0 = 1/2 $ is the variance in $q$ quadrature of the vacuum state and $(\Delta q_{\text{obs}})^2$ is the observed variance in $q$ quadrature of the measured signal state. As the value of $\xi$ is normalized with respect to the vacuum variance, the channel excess noise $\xi$ is reported in the shot noise units (SNU) and independent of different conventions of defining quadrature operators. 

Apart from the shot noise, there are several contributions to the total noise in the measurement data such as preparation noise, detector noise and noises introduced in the fiber due to Raman scattering. As we treat the detection noise as trusted, we assume all other contributions are under Eve's control. In other words, all additional noises beyond the shot noise except for the detector noise become a part of the effective quantum channel regardless of the physical origin of each noise component, and they contribute to the value of the excess noise $\xi$. In the literature, the value of the excess noise $\xi$ is commonly reported at the input of the quantum channel corresponding to measuring $(\Delta q_{\text{obs}})^2$ at the output of Alice's lab. By choosing this convention of reporting the value of excess noise, we may alternatively imagine that this effective quantum channel first introduces the amount of excess noise $\xi$ to the signal state at the input of the channel and the rest of this quantum channel is then lossy but noise-free. Under this channel model, a coherent state $\ket{\alpha}$, after transmitting through this quantum channel, becomes a displaced thermal state centered at $\sqrt{\eta_t} \alpha$ with its variance $\frac{1}{2}(1+ \eta_t \xi)$ for each quadrature.

\subsection{Simulated statistics}
From our simulation, the simulated state $\sigma^x_{B}$ conditioned on the choice of $x$ is a displaced thermal state whose Wigner function is 
\begin{aeq}
W_{\sigma^x_{B}}(\gamma)  &=  \frac{1}{\pi} \frac{1}{\frac{1}{2}(1+\eta_t\xi)}\exp[-\frac{\abs{\gamma-\sqrt{\eta_t}\alpha_x}^2}{\frac{1}{2}(1+\eta_t\xi)}].
\end{aeq}

When Bob applies his heterodyne measurement described by the POVM $\{G_y\}$, the probability density function $P(y|x)$ for the measurement outcome $y$ conditioned on Alice's choice $x$ is 
\begin{aeq}\label{eq:pdf_simplified}
P(y|x)
&=\frac{1}{\pi (1+\frac{1}{2}\eta_d\eta_t \xi + \nu_{\text{el}})} \exp[-\frac{\abs{y-\sqrt{\eta_d\eta_t}\alpha_x}^2}{1+\frac{1}{2}\eta_d\eta_t \xi + \nu_{\text{el}} }].
\end{aeq}The observables defined in Eq. (\ref{eq:observables}) have the following expectation values from the simulation: 
\begin{aeq}\label{eq:simulatedStatistics}
\langle \hat{F}_{Q}  \rangle_x &=  \sqrt{2\eta_d\eta_t} \Re(\alpha_x), \\
\langle \hat{F}_{P}  \rangle_x &=\sqrt{2\eta_d\eta_t} \Im(\alpha_x), \\
\langle \hat{S}_{Q}  \rangle_x 
&=2\eta_d\eta_t\Re(\alpha_x)^2 +1+\frac{1}{2}\eta_d\eta_t \xi + \nu_{\text{el}},\\
\langle \hat{S}_{P}  \rangle_x 
&= 2\eta_d\eta_t\Im(\alpha_x)^2 + 1+\frac{1}{2}\eta_d\eta_t \xi + \nu_{\text{el}}.
\end{aeq}

\subsection{Estimation of error correction cost}
We estimate the cost of error correction from the simulated statistics. From the probability density function $P(y|x)$ shown in Eq. (\ref{eq:pdf_simplified}), we can obtain the joint probability distribution $\widetilde{P}(x, z)$ for Alice's choice $\mathbf{X}=x$ and Bob's discretized key value $\mathbf{Z}=z$ by the following integral
\begin{aeq}
\widetilde{P}(z|x) = \int_{\Delta_a}^{\infty}dr \; r \int_{\frac{2z-1}{4}\pi}^{\frac{2z+1}{4}\pi} d\theta P(re^{i\theta}|x).
\end{aeq}

Since $\widetilde{P}(x)=p_x = \frac{1}{4}$, we then obtain the joint probability distribution $\widetilde{P}(x,z)= \widetilde{P}(z|x) \widetilde{P}(x)$. Using the definition of $\I(\mathbf{X};\mathbf{Z})$ in terms of $\widetilde{P}(x,z)$, we can approximate the cost of error correction by Eq. (\ref{eq:delta_ec_RR}) for the reverse reconciliation scheme considered in this work. When $\Delta_a$ is not zero, that is, in the presence of postselection, the sifting factor $p_{\text{pass}}$ is the sum of $\widetilde{P}(x,z)$ over $x, z \in \{0,1,2,3\}$. We then renormalize the probability distribution before plugging it in the definition of $\I(\mathbf{X};\mathbf{Z})$.

For the purpose of illustration, we choose the error correction efficiency $\beta$ to be 95\% for our simulations, which is around typical values for the state-of-the-art error correction codes (see e.g.  Ref. \cite{Milicevic2018}).

\section{Key rate in the absence of postselection}\label{sec:results_nops}
In this section, we present results when no postselection is performed, that is, $\Delta_a=0$.  We make two comparisons. The first one is to compare key rates in the trusted and untrusted detector noise scenarios. The second one is to analyze how different imperfections in detectors affect key rates in the trusted detector noise scenario. 

\subsection{Comparison between trusted and untrusted detector noise scenarios}
\begin{figure}[h]
\includegraphics[width = \linewidth]{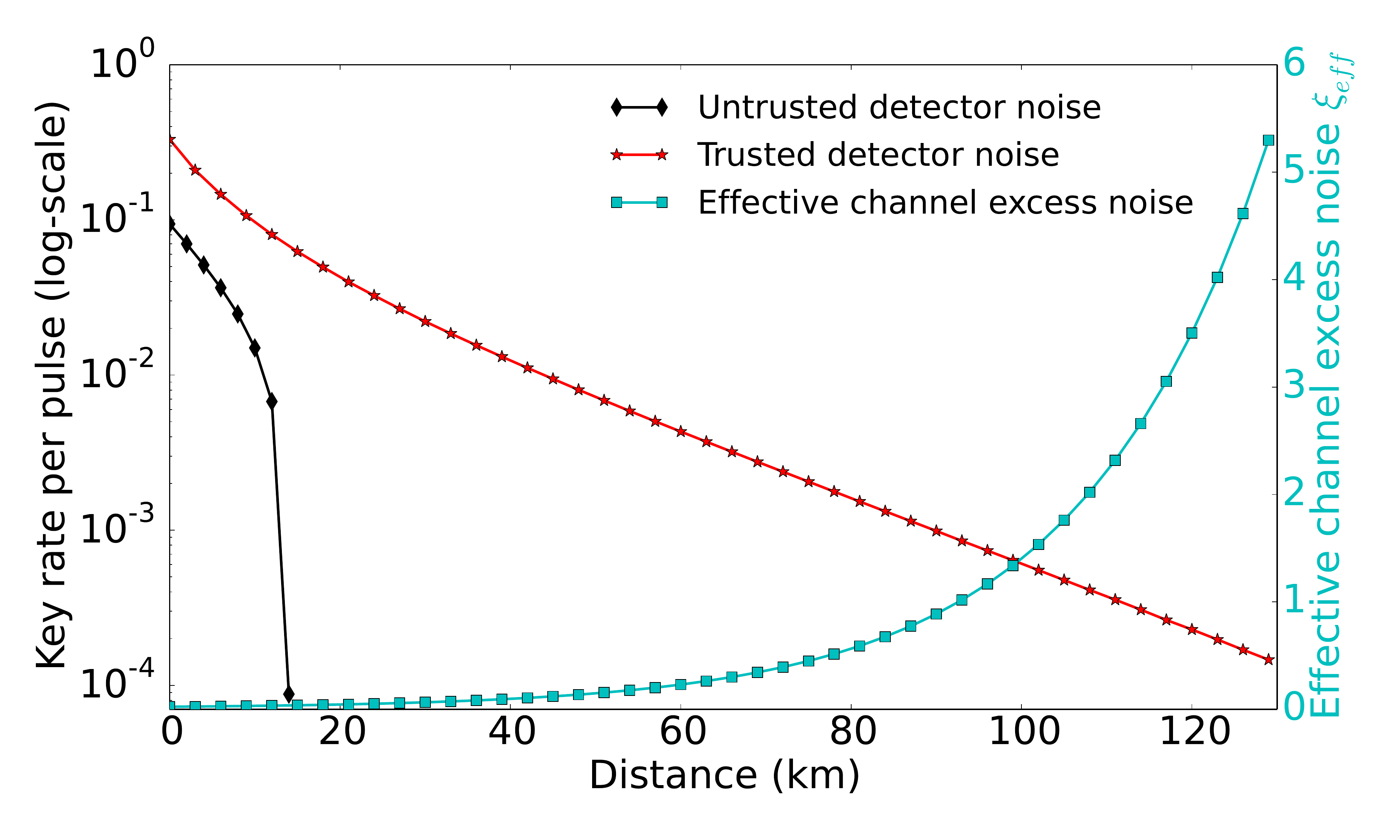}
\caption{\label{fig:RRtype1comp} Secure key rate versus the transmission distance for untrusted detector noise (black diamonds) and trusted detector noise (red stars) scenarios. The excess noise is $\xi = 0.01$ at the input of the quantum channel. Parameters for detector are $\eta_d = 0.719$, $\nu_{\text{el}} = 0.01$\cite{Soh2015}. The error correction efficiency is $\beta = 0.95$. The coherent state amplitude is optimized via a coarse-grained search over the interval $[0.5,0.9]$ with a step size of $0.05$ and the channel transmittance is $\eta_t = 10^{-0.02 L}$ for each distance $L$ in kilometers. The effective channel excess noise in the untrusted detector scenario is shown with the $y$ axis on the right. At 20 km, the effective channel excess noise $\xi_{\text{eff}}$ is roughly 0.045.}
\end{figure}

For this comparison, we supply the same set of simulated data from Eq. (\ref{eq:simulatedStatistics}) to the optimization problem for the untrusted detector noise scenario in Eq. (\ref{eq:optimization}) and the one for the trusted detector noise scenario in Eq. (\ref{eq:optimization_reformulated}). For simulation, we choose parameters $\eta_d = 0.719$, $\nu_{\text{el}} = 0.01$ from Ref.  \cite{Soh2015} for illustration. The result is shown in Fig. \ref{fig:RRtype1comp}. 

As we can see from this figure, the key rate of the untrusted detector noise scenario drops quickly at a short distance less than 20 km even though the electronic noise is only 0.01 SNU, which is a low value compared to detectors used in many other CV experiments. On the other hand, the key rate in the trusted detector noise scenario extends to much longer distances, which exhibits a similar behavior as the results shown in Ref. \cite{Lin2019} when the detector is treated as ideal. One explanation for this behavior is that in Ref. \cite{Lin2019}, we observe that the key rate for the QPSK scheme drops quickly when the channel excess noise $\xi$ is large. Since the value of $\xi$ is reported at the input of the quantum channel while the value of $\nu_{\text{el}}$ is measured at Bob's side, to treat $\nu_{\text{el}}$ as a part of channel excess noise in the untrusted detector noise scenario, one needs to define the effective value of $\xi$ to include the value of $\nu_{\text{el}}$. For the effective value $\xi_{\text{eff}}$, the electronic noise $\nu_{\text{el}}$ needs to be scaled by a factor of $1/\eta_t$ (in addition to $1/\eta_d$), which is large for slightly long distances as $\eta_t$ becomes small. As a result, the redefined value $\xi_{\text{eff}}$ of $\xi$ is quite large as shown in Fig. \ref{fig:RRtype1comp} and this behavior of key rate is then expected. By the observation made from this figure, it is not surprised that for a larger value of electronic noise, the key rate in the untrusted detector noise scenario would drop to zero at an even shorter distance.

\subsection{Detector imperfection in the trusted detector noise scenario}
\begin{figure}[h]
\includegraphics[width = \linewidth]{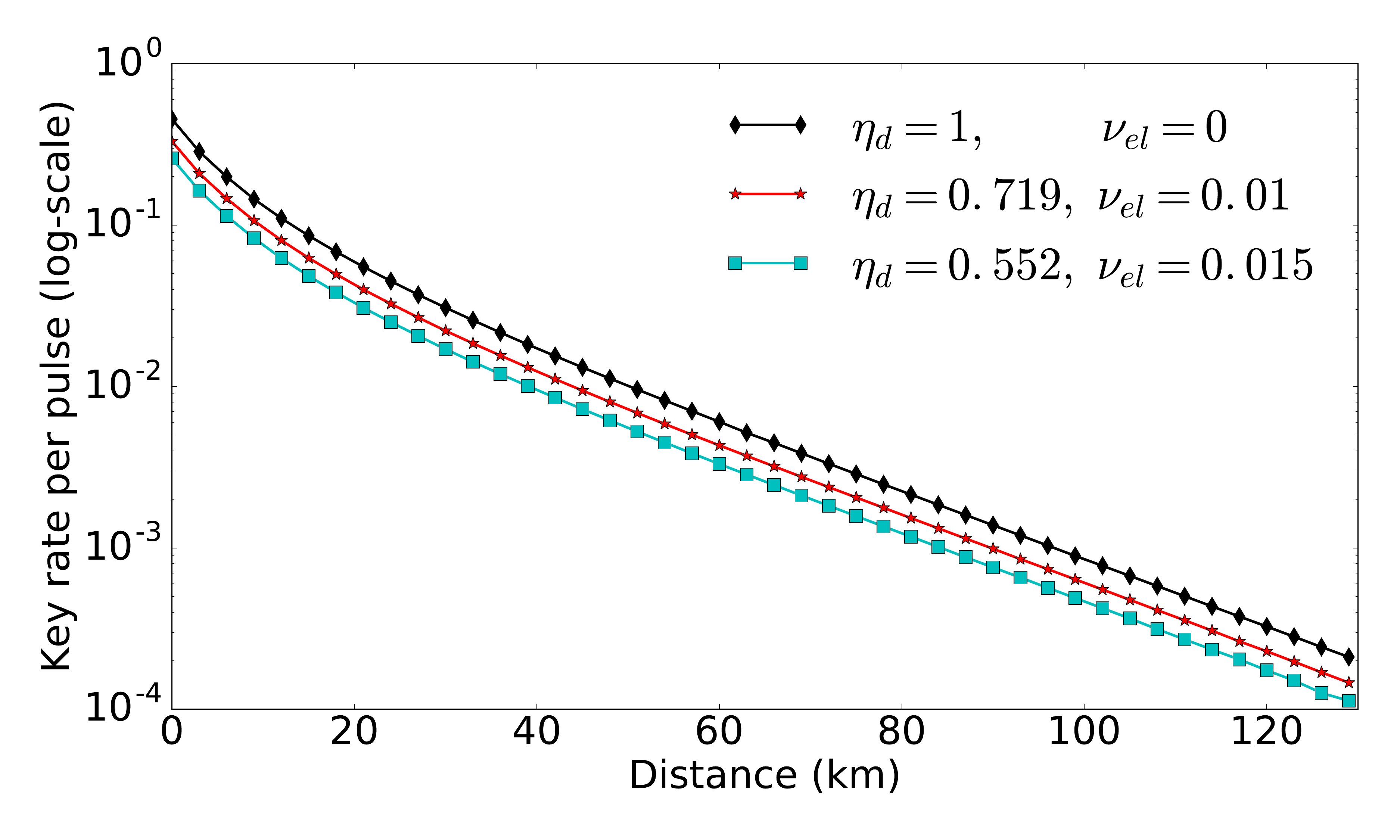}
\caption{\label{fig:type2comp1} Secure key rate versus transmission distance for different detector imperfections reported in experiments in a comparison to the ideal detector. Other parameters are the excess noise $\xi = 0.01$, error-correction efficiency $\beta= 0.95$,  and the transmittance $\eta_t = 10^{-0.02 L}$ for each distance $L$ in kilometers. For each distance, the coherent state amplitude $\alpha$ is optimized via a coarse-grained search in the interval $[0.5,0.9]$ with a step size of $0.05$. Black curve with diamond markers is for the ideal heterodyne detector; red curve with star markers is for the detector used in Ref. \cite{Soh2015}; cyan curve with square markers is for the detector used in Ref. \cite{Jouguet2013}.}
\end{figure}

\begin{figure}[h]
\subfloat[]{\label{fig:type2comp2a}\includegraphics[width = \linewidth]{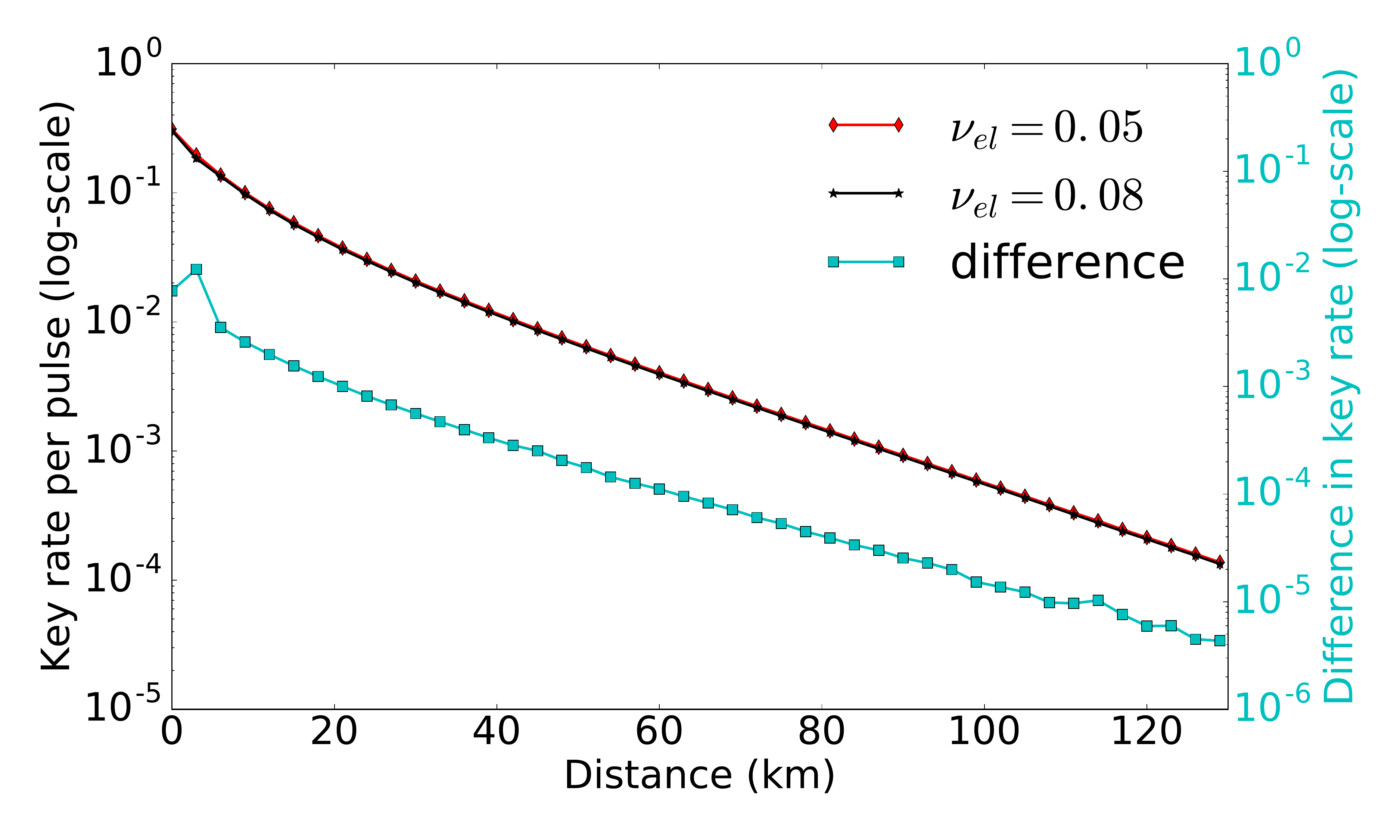}}\\
\subfloat[]{\label{fig:type2comp2b}\includegraphics[width = \linewidth]{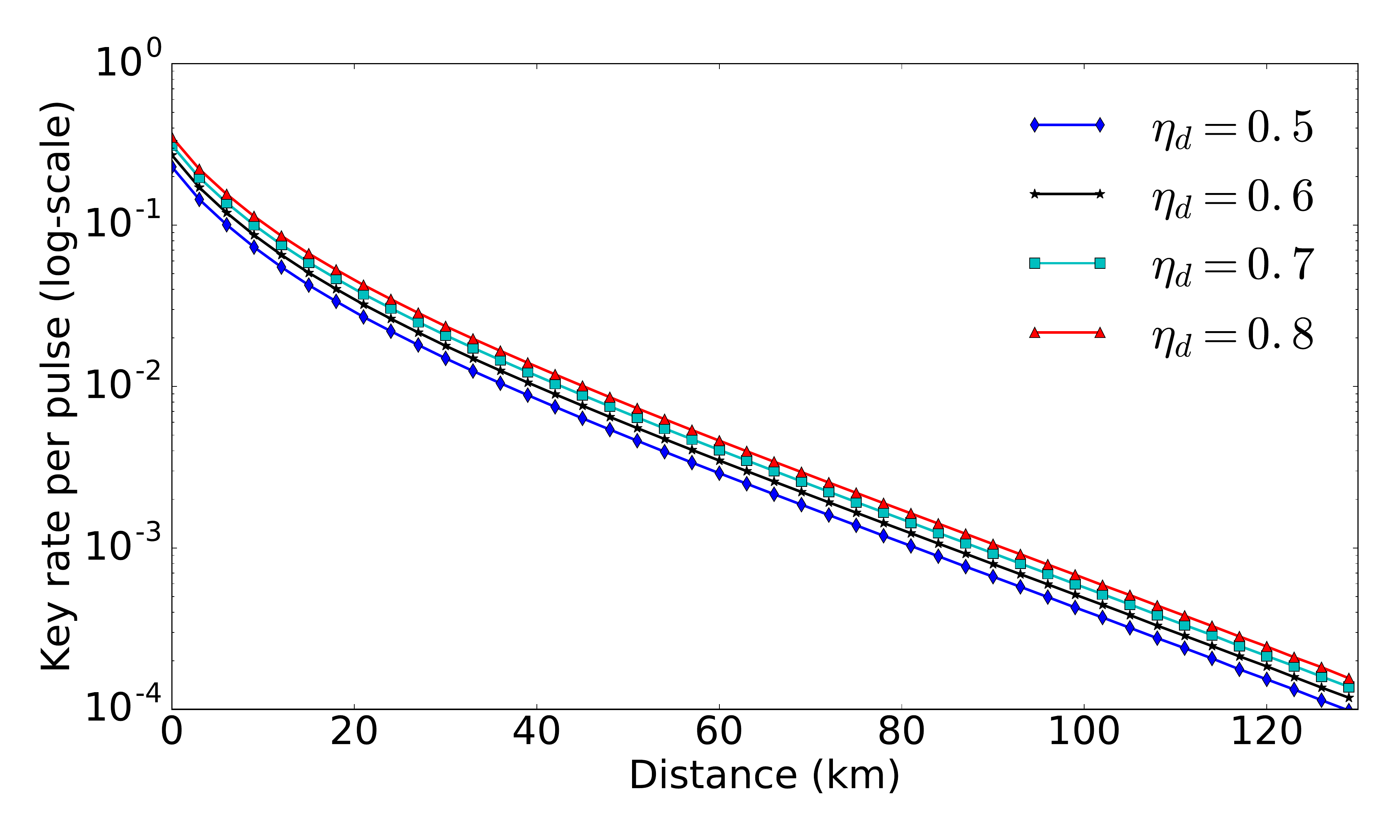}}
\caption{\label{fig:type2comp2} Secure key rate versus transmission distance for different detector imperfections with the excess noise $\xi = 0.01$. For both plots, the coherent state amplitude is optimized via a coarse-grained search over the interval $[0.5,0.9]$ with a step size 0.05 and $\beta = 0.95$. (a) Comparison of key rates between two values of the electronic noise when the detector efficiency is set to be $\eta_d=0.7$ for both curves. The difference of two curves is also plotted with the secondary $y$-axis on the right. (b) Comparison of key rates for different values of detector efficiency when the electronic noise is $\nu_{\text{el}} =0.05.$ }
\end{figure}

To guide the experimental implementation of the QPSK scheme, we may be interested in the robustness of the protocol in the presence of detector inefficiency and electronic noise in the trusted detector noise scenario. For this purpose, we investigate the effects of different levels of detector efficiency and electronic noise on the key rate. For curves in Figs. \ref{fig:type2comp1} and \ref{fig:type2comp2}, our simulation uses the same channel model but different detector imperfections, that is, in Eq. (\ref{eq:simulatedStatistics}), the same values of channel parameters $\eta_t$ and $\xi$ but different values of detector efficiency $\eta_d$ and electronic noise $\nu_{\text{el}}$ (as specified in the captions) for different curves. 

In Fig. \ref{fig:type2comp1}, we choose values of $\eta_d$ and $\nu_{\text{el}}$ for a homodyne detector from two experiments \cite{Jouguet2013, Soh2015} and compare these results with the ideal detector. For the comparison, we optimize $\alpha$ via a coarse-grained search for each distance. We see that with a noisy heterodyne detector, the key rate drops moderately from the key rate of using an ideal detector. The amount of decrease is like a constant prefactor in the key rate. As the detector is noisier, the key rate becomes lower as expected.  

To show that different values of electronic noise have little impacts on the secure key rates in the trusted noise scenario, we compare key rates with two choices of the electronic noise value in Fig. \ref{fig:type2comp2a} while we fix the value of detector efficiency $\eta_d$ to be 0.7. As the key rate difference is relatively small between the curve with $\nu_{\text{el}}=0.05$ and that with $\nu_{\text{el}}=0.08$, we also plot the difference of key rate (that is, the key rate with $\nu_{\text{el}}=0.05$ minus the key rate with $\nu_{\text{el}}=0.08$) in the same figure. (Note that the non-smoothness in the curve of difference is due to the coarse-grained search for the coherent state amplitude in the presence of the numerical performance issue discussed in Ref. \cite{Lin2019}.) We observe that when the electronic noise is trusted, its impact on the secure key rates is insignificant. This result eases the requirements of a detector in a CVQKD experiment with the QPSK scheme. Similarly, we investigate the effects of detector efficiency in Fig. \ref{fig:type2comp2b}. In particular, we fix the value of electronic noise $\nu_{\text{el}}$ to be 0.05 SNU and plot four choices of detector efficiency between 0.5 and 0.8. We see the key rate curves are close to each other. 

\begin{figure}[h]
\includegraphics[width = \linewidth]{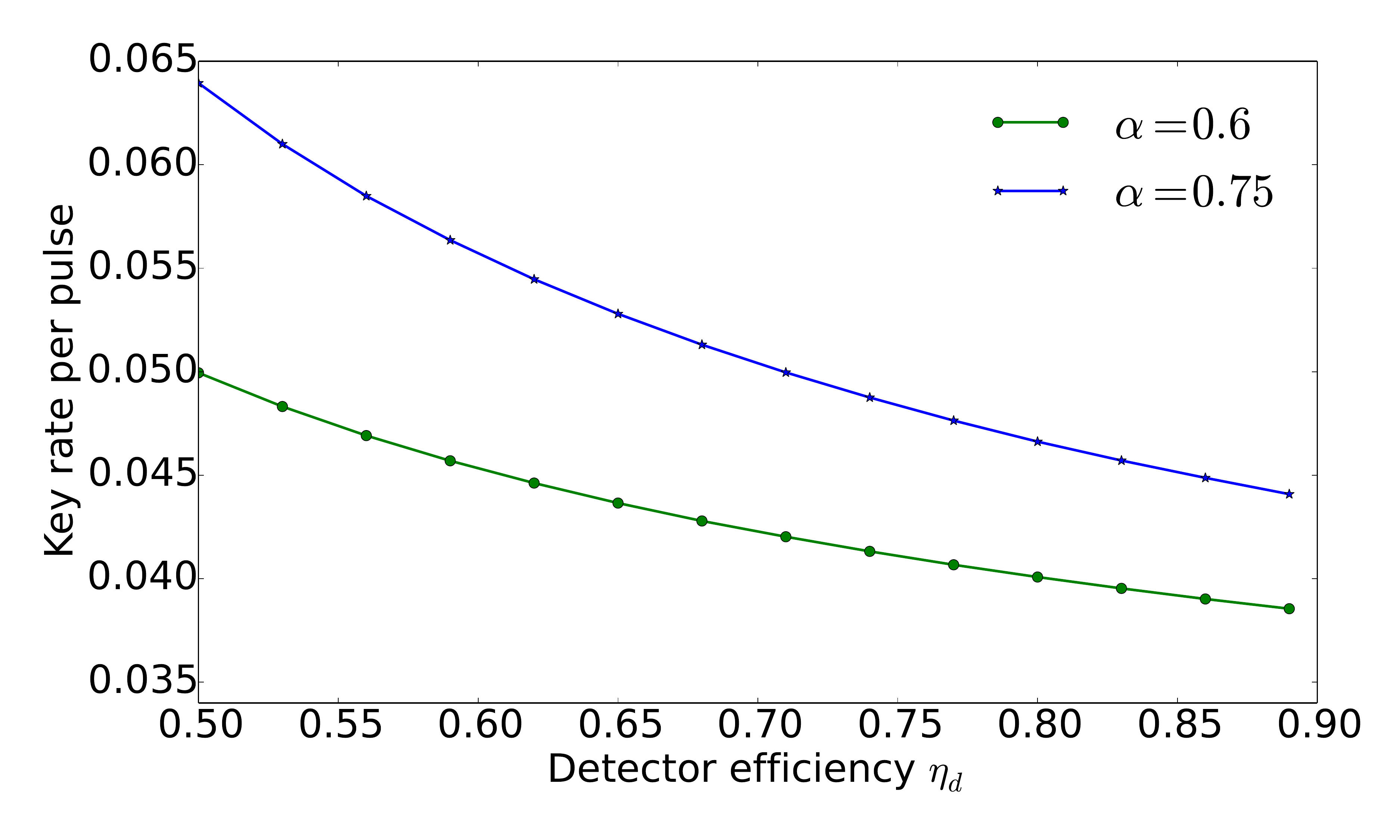}
\caption{\label{fig:type2comp3} Secure key rate versus the detector efficiency $\eta_d$ for a fixed value of total transmittance $\eta :=\eta_t \eta_d = 0.3155$. This figure studies the tradeoff between the key rate and the amount of trusted loss. Other parameters are the excess noise $\xi = 0.01$, the electronic noise $\nu_{\text{el}} = 0.01$,  and the error-correction efficiency $\beta = 0.95$. We include two curves for different choices of coherent state amplitude $\alpha$.}
\end{figure}

In Fig. \ref{fig:type2comp3}, we investigate the tradeoff between trusting the detector efficiency and lumping it together with the channel transmittance, similar to a scenario studied in Ref. \cite{Zhang2020} for discrete-variable systems. For the fixed amount of total transmittance $\eta := \eta_t\eta_d$, it is interesting to see how trusting different values of detector efficiency affects the key rate. We observe that when the value of the product of channel transmittance $\eta_t$ and detector efficiency $\eta_d$ is fixed, if the detector efficiency $\eta_d$ is lower, meaning that if more contribution to the total transmittance $\eta$ is trusted, then the key rate is higher. This observation is similar to the observation made for discrete-variable systems in Ref. \cite{Zhang2020}.  

To summarize, in a discrete modulation experiment, if one is able to obtain accurate values of $\eta_d$ and $\nu_{\text{el}}$ by a suitable calibration procedure and able to maintain a low level of the effective channel excess noise $\xi$ to a value like $0.01$, then the QPSK scheme is able to extend to a distance beyond 100 km in the asymptotic regime. We remark that the optimal amplitude for the QPSK scheme in the trusted detector noise scenario is around 0.75 corresponding to a mean photon number of around 0.56, similar to the optimal amplitude in the ideal or untrusted detector noise scenario reported in our previous work \cite{Lin2019}. This mean photon number is much lower than that for Gaussian modulation schemes. 
\section{Key rate with postselection}\label{sec:results_ps}
In this section, we investigate the effects of postselection in the trusted detector noise scenario. As demonstrated in our previous analysis \cite{Lin2019}, postselection of data can improve the key rate of the QPSK scheme in the untrusted detector noise scenario. Postselection is simple to implement in an experiment. It not only improves the key rate but also reduces the required volume of data postprocessing. Thus, it is advantageous to include a postselection step in the protocol. As expected, we show here that this advantage also exists in the trusted detector noise scenario.

\begin{figure}[h]
\subfloat[]{\label{fig:ps_L50}\includegraphics[width = 0.5\linewidth]{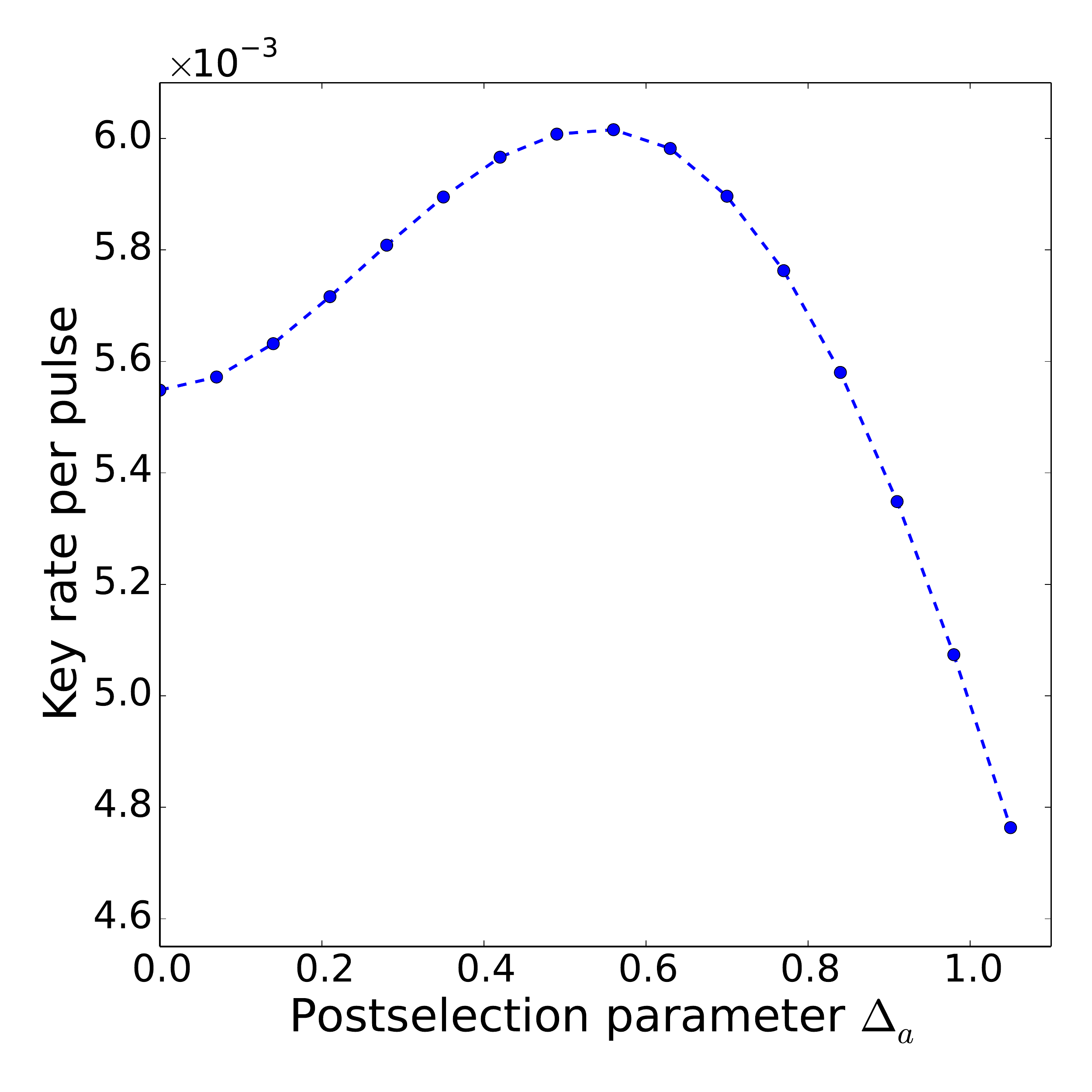}}
\subfloat[]{\label{fig:ps_L75}\includegraphics[width = 0.5\linewidth]{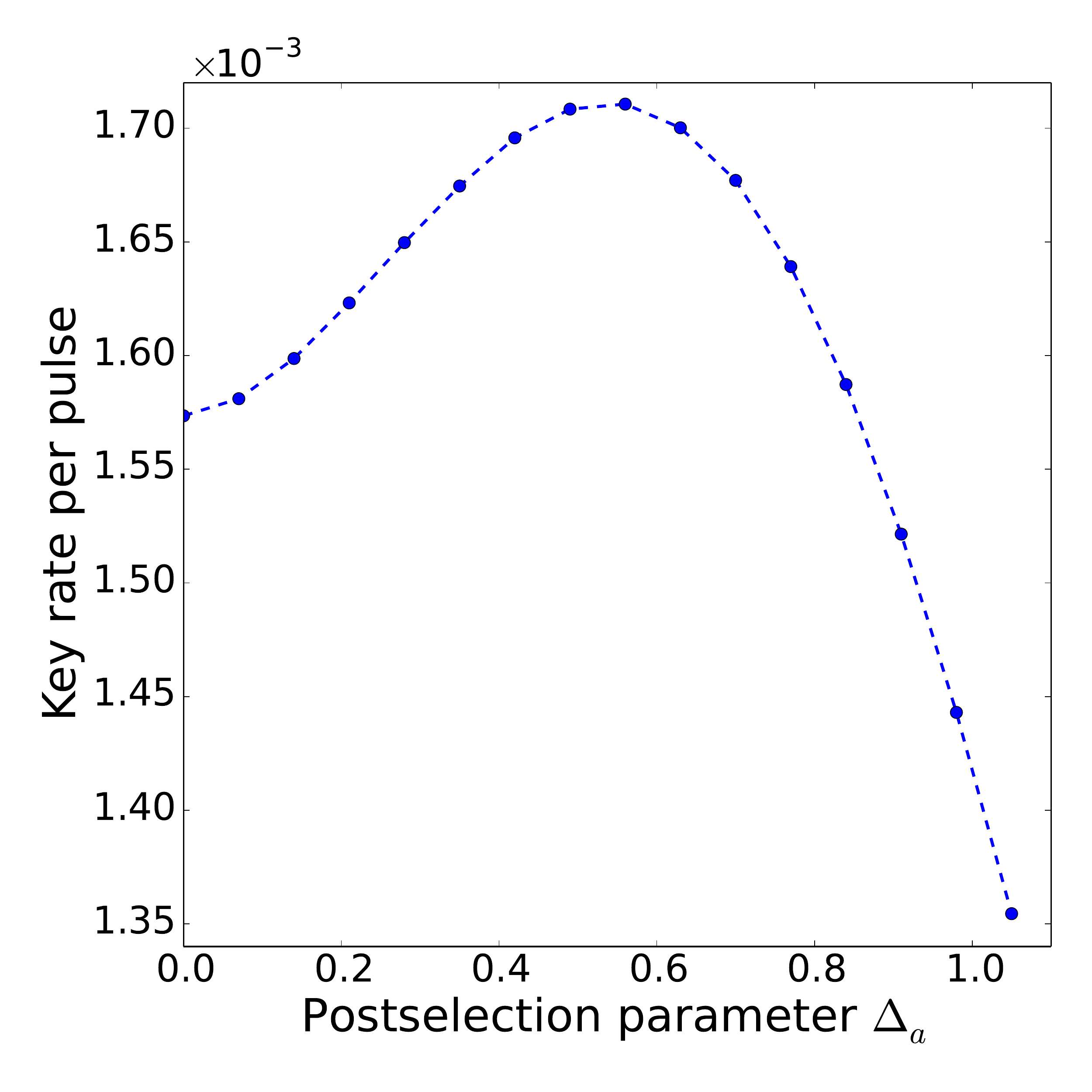}}
\caption{\label{fig:ps_fixed_distance} (a) Secure key rate versus postselection parameter $\Delta_a$ for $L=50$ km. (b) Secure key rate versus postselection parameter $\Delta_a$ for $L=75$ km. For both plots, the channel excess noise is $\xi=0.01$ and the error-correction efficiency $\beta = 0.95$. The coherent state amplitude is optimized via a coarse-grained search in the interval [0.6, 0.8] with a step size of 0.05. Parameters for detectors are $\eta_d = 0.552$ and $\nu_{\text{el}} = 0.015$ from Ref. \cite{Jouguet2013}.}
\end{figure}

In Fig. \ref{fig:ps_fixed_distance}, we search for the optimal postselection parameter for different transmission distances and take the distances $L = 50$ km and  $L=75$ km as examples. For this figure, we also optimize the choice of coherent state amplitude via a coarse-grained search. The $x$ axis in each plot is the postselection parameter $\Delta_a$. We observe the optimal value of the postselection parameter $\Delta_a$ is around 0.6 for both $L= 50$ km and $L=75$ km. We also observe that the optimal choice of the postselection parameter $\Delta_a$ does not change significantly for different distances.

\begin{figure}[h]
\includegraphics[width = \linewidth]{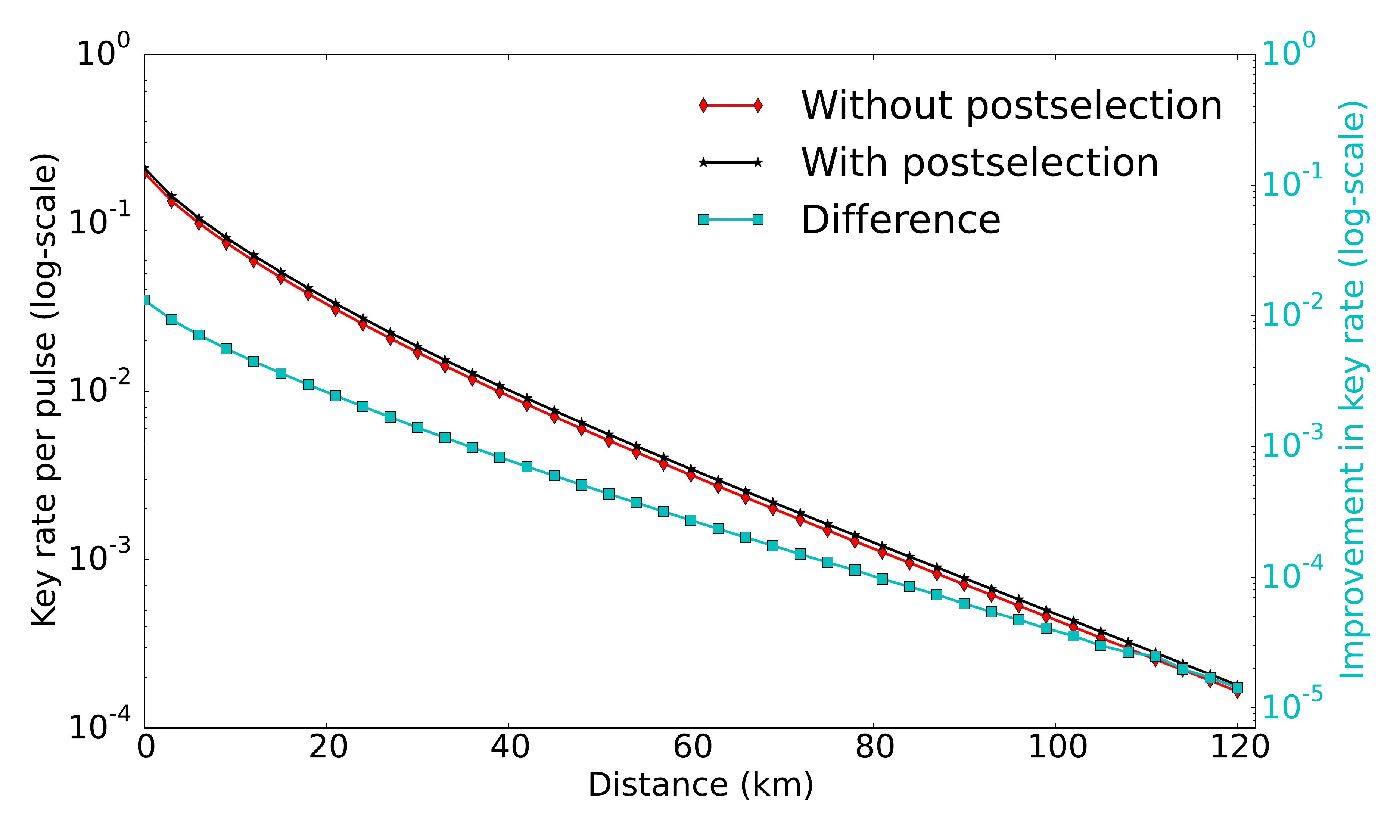}
\caption{\label{fig:ps_optimized} Comparison of key rates with or without postselection. Detector parameters are from Ref. \cite{Jouguet2013} where $\eta_d = 0.552$ and $\nu_{\text{el}}=0.015$. The difference of two curves is also plotted with the secondary $y$ axis on the right. Other parameters are the channel excess noise $\xi=0.01$, the coherent state amplitude $\alpha = 0.75$,  and the error-correction efficiency $\beta = 0.95$. The postselection parameter is optimized via a coarse-grained search in the interval [0.45,0.7] with a step size 0.05. }
\end{figure}

In Fig. \ref{fig:ps_optimized}, we show the key rate as a function of transmission distance for two scenarios: with or without postselection. Since the optimal postselection parameter does not change significantly for different distances, we optimize the postselection parameter $\Delta_a$ via a coarse-grained search in a restricted interval. For this figure, we fix the coherent state amplitude to be 0.75 and the channel excess noise $\xi$ to be 0.01. We see postselection can indeed improve the key rate. The percentage of improvement compared to the key rate without postselection is roughly between 5\% to 8\% and the probability of being postselected is around 70\% to 80\%. Thus, postselection can reduce the amount of data for postprocessing by around 20\% to 30\% while improving the key rate. 

We end this section with a remark on the postselection pattern. The postselection pattern (see Fig. \ref{fig:keymap}) studied in this work is a simple, intuitive,  and convenient choice when we evaluate the region operators. However, it is not necessarily the optimal way to postselect data  \cite{Silberhorn2002, Heid2006}. It is an interesting future work to investigate other patterns of postselection.

\section{Summary and future directions}\label{sec:outlook}
We provide a method to analyze the asymptotic security of a discrete modulation scheme of CVQKD in the trusted detector noise scenario where both nonunity detector efficiency and electronic noise are trusted. In particular, we find the POVM elements corresponding to a noisy heterodyne detector. As we demonstrate our method on the quadrature phase-shift keying scheme, we show that when the detector imperfection is trusted, the key rates are similar to the one with the ideal heterodyne detector studied previously \cite{Lin2019}. Our analysis in this work eases the requirements of an experimental implementation of the discrete modulation scheme as the detector imperfection is usually a major source of noise. 

We point out the limitations in the current work. First, the analysis in this work is still restricted to the asymptotic scenario. We notice that there is a recent work on the finite key analysis of binary modulation protocol \cite{Matsuura2020}. However, the key rate there was very pessimistic and one expects that quadrature-shift keying schemes will have much better performance. It remains an open question to provide a finite key analysis of general discrete modulation beyond binary modulation. As we recently extend the underlying numerical method used in this security analysis to finite-key regime \cite{George2020}, we hope to perform the finite key analysis for discrete modulation schemes, especially the protocol studied in this work. However, there remain technical challenges to solve before such an analysis can be carried out and thus we leave the finite key analysis for future work.  The second limitation is the same photon-number cutoff assumption used in Refs. \cite{Ghorai2019, Lin2019}. While numerical evidences show that our results are stable when the cutoff photon number is chosen appropriately, we plan to have a more rigorous analysis on the effects of truncation beyond numerical evidences in future work. Thirdly, we present simulation results in a simple scenario where two homodyne components are treated as identical. This scenario is commonly assumed in previous studies of Gaussian modulation schemes. In the simple scenario, we are able to provide simplified expressions for region operators and observables used in the key rate optimization problem. However, our principles presented in this paper work for the general case where two detectors are not identical. To handle the general case, one may perform the numerical integration of POVM element $G_y$'s to find necessary operators in the photon-number basis from the photon-number basis representation of each POVM element $G_y$ shown in Appendix \ref{app_sec:general_case}. It may become numerically demanding to perform these integrals. Alternatively, one may attempt to simplify expressions analytically similar to what we have done for the simple case. It remains as a technical question to efficiently compute the matrix elements of operators defined in terms of $G_y$ in the photon-number basis, which we expect can be solved. Nevertheless, this current limitation does not affect the principles and methodology we present in this work about the treatment of trusted detector noise. It is also expected that observations in the general case will be similar to observations we make here in the simple case.

Finally, we remark on the generality of our method of treating trusted detector noise. If a different physical model of a detector is adopted (which needs to be verified experimentally), we expect that a similar method as described here can be used to find a correct POVM description for the given physical model and then this POVM can be used in the security analysis.

\begin{acknowledgments}
We thank Mi Zou and Feihu Xu for helpful discussions related to experiments. We also thank Twesh Upadhyaya for code review. The work is performed at the Institute for Quantum Computing (IQC), University of Waterloo, which is supported by Industry Canada. J. L. acknowledges the support of Mike and Ophelia Lazaridis Fellowship from IQC. The research has been supported by NSERC under the Discovery Grants Program, Grant No. 341495, and under the Collaborative Research and Development Program, Grant No. CRDP J 522308-17. Financial support for this work has been partially provided by Huawei Technologies Canada Co., Ltd.
\end{acknowledgments}

\appendix

\section{Derivation of noisy heterodyne detection POVM via Wigner functions}\label{app:povm}

\subsection{Basic Wigner functions}
As we use the Wigner function approach for our derivation, we recall useful expressions from Ref. \cite{Leonhardt2010} for later references.

To calculate $\Tr(FG)$ for two operators $F$ and $G$ in terms of their Wigner functions $W_F$ and $W_G$, the overlap formula is 
\begin{aeq}\label{eq:overlap}
\Tr(FG) = \pi \int d^2 \alpha \; W_F(\alpha)W_G(\alpha).
\end{aeq}We can easily generalize the formula to multimode cases. The input-output Wigner functions under a beam-splitter transformation whose transmittance is $\eta$ are related by
\begin{aeq}\label{eq:wigner_beamsplitter}
W_{\text{out}}(\alpha, \beta) = W_{\text{in}}(\sqrt{\eta}\alpha + \sqrt{1-\eta}\beta, \sqrt{1-\eta}\alpha - \sqrt{\eta}\beta).
\end{aeq}

We list Wigner functions for some quantum states that are relevant for our discussions here. The Wigner function of a vacuum state $\ket{0}$ is 
\begin{aeq}\label{eq:wigner_vacuum}
W_{\ket{0}}(\gamma) = \frac{2}{\pi} e^{-2\abs{\gamma}^2}.
\end{aeq}
The Wigner function of a thermal state $\rho_{\text{th}}(\bar{n})$ with the mean photon number $\bar{n}$ is
\begin{aeq}\label{eq:wigner_thermal}
W_{\rho_{\text{th}}(\bar{n})}(\gamma) = \frac{2}{\pi}\frac{1}{1+2\bar{n}} e^{-\frac{2\abs{\gamma}^2}{1+2\bar{n}}}.
\end{aeq}
The Wigner function of a displaced thermal state (DTS) $\rho_{\text{DTS}}(\alpha,\bar{n}):=\hat{D}(\alpha)\rho_{\text{th}}(\bar{n})\hat{D}^{\dagger}(\alpha)$ with the amount of displacement $\alpha$ is
\begin{aeq}\label{eq:wigner_displaced_thermal}
W_{\rho_{\text{DTS}}(\alpha,\bar{n})}(\gamma) = \frac{2}{\pi}\frac{1}{1+2\bar{n}} e^{-\frac{2\abs{\gamma-\alpha}^2}{1+2\bar{n}}}.
\end{aeq}We notice that if we set $\alpha=0$, it reduces to Eq. (\ref{eq:wigner_thermal}).

It is also useful to note the Wigner functions of a squeezed thermal state (STS) and of a displaced squeezed thermal state (DSTS). Let $\hat{S}(\xi)$ denote the squeezing operator with a squeezing parameter $\xi$. For our discussion, we restrict $\xi \in \mathbb{R}$. For a squeezed thermal state $\rho_{\text{STS}}(\xi,\bar{n}):=\hat{S}(\xi)\rho_{\text{th}}(\bar{n})\hat{S}^{\dagger}(\xi)$, its Wigner function reads (see, e.g., Eq. (4.13) of Ref. \cite{Kim1989})
\begin{widetext}
\begin{aeq}
W_{\rho_{\text{STS}}(\xi,\bar{n})}(\gamma)=\frac{2}{\pi}\frac{1}{1+2\bar{n}}\exp{-2 [\frac{e^{2\xi}\Re(\gamma)^2+e^{-2\xi}\Im(\gamma)^2}{1+2\bar{n}}]}
\end{aeq}

The Wigner function of a displaced squeezed thermal state $\rho_{\text{DSTS}}(\alpha,\xi,\bar{n}):=\hat{D}(\alpha)\hat{S}(\xi)\rho_{\text{th}}(\bar{n})\hat{S}^{\dagger}(\xi)\hat{D}^{\dagger}(\alpha)$ can be similarly written as 
\begin{aeq}\label{eq:wigner_dsts}
W_{\rho_{\text{DSTS}}(\alpha,\xi,\bar{n})}(\gamma)=\frac{2}{\pi}\frac{1}{1+2\bar{n}}\exp{-2 [\frac{e^{2\xi}\Re(\gamma-\alpha)^2+e^{-2\xi}\Im(\gamma-\alpha)^2}{1+2\bar{n}}]}
\end{aeq}
\end{widetext}

\subsection{Derivation}
\begin{figure}[h]
\includegraphics[width=\linewidth]{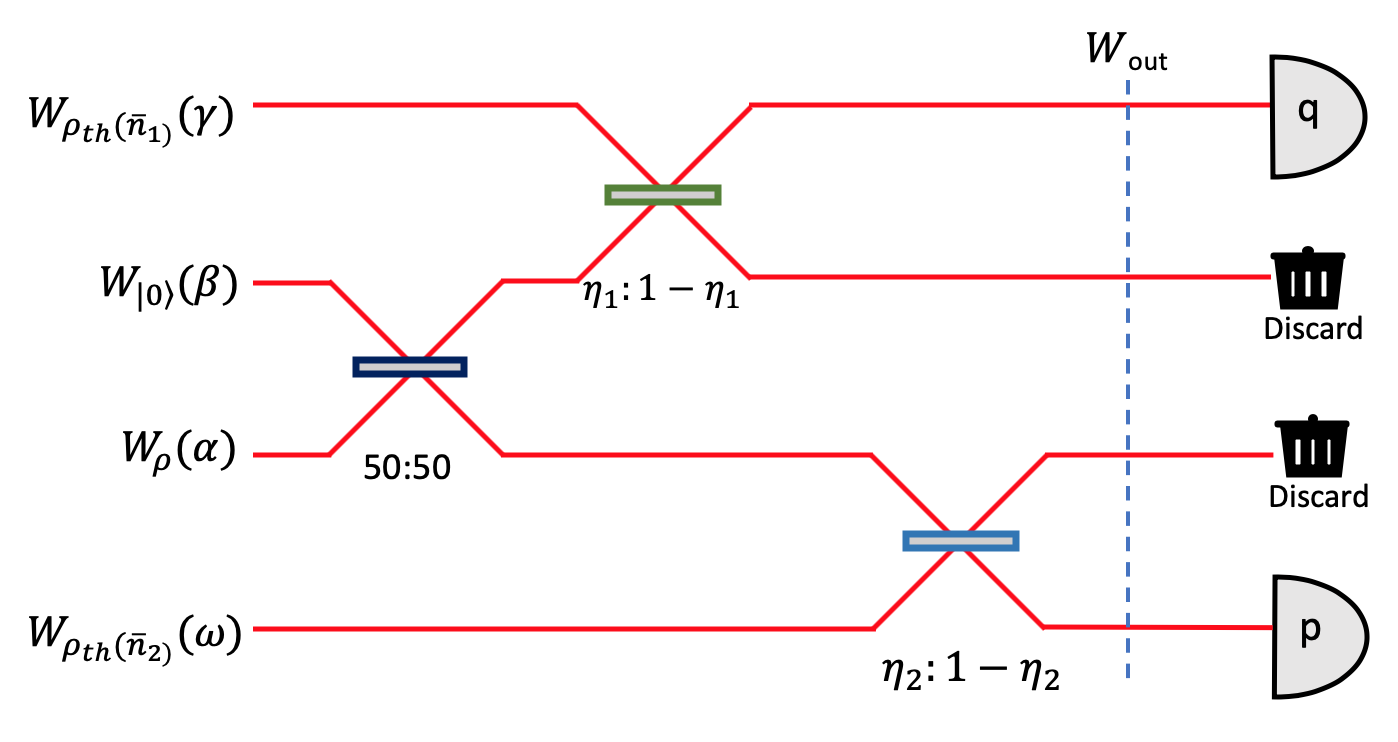}
\caption{\label{fig:noisy_het_eq} A concise but equivalent view of the noisy heterodyne detector model depicited in Fig. \ref{fig:noisy_het}. Input modes are labeled in terms of Wigner functions.}
\end{figure}
As the physical model of a noisy heterodyne detector is presented in Fig. \ref{fig:noisy_het}, our goal here is to find the corresponding POVM elements that correctly produce the probability density function $P(y)$ of obtaining an outcome $y \in \mathbb{C}$ for an arbitrary input state $\rho$ to the detector. In our trusted noise model, the homodyne detector for the $q$ quadrature measurement has its detector efficiency $\eta_1$ and electronic noise $\nu_{1}$ which is related to a thermal state of the mean photon number $\bar{n}_1 =\frac{\nu_{1}}{2(1-\eta_1)}$. Similarly, the homodyne detector for the $p$ quadrature measurement has its detector efficiency $\eta_2$ and electronic noise $\nu_{2}$ which corresponds to a thermal state with the mean photon number $\bar{n}_2 =\frac{v_{2}}{2(1-\eta_2)}.$ Figure \ref{fig:noisy_het_eq} shows a compact but equivalent representation of Fig. \ref{fig:noisy_het} with Wigner functions associated to input modes. In this setup, for an output state $W_{\text{out}}(\alpha,\beta,\gamma,\omega)$ at the step labeled in Fig. \ref{fig:noisy_het_eq}, we measure the $q$ quadrature of the mode $\alpha$ and $p$ quadrature of the mode $\beta$ with two ideal homodyne detectors, and discard the rest modes $\gamma$ and $\omega$. The Wigner function of an ideal homodyne detector for the $q$ quadrature measurement that produces a measurement outcome $\Re(y)$ is $W_{H_{\Re(y)}}(\alpha) = \frac{1}{\sqrt{2}\pi}\delta(\Re(\alpha)-\frac{\Re(y)}{\sqrt{2}})$ where $\delta$ is the Dirac delta function and similarly, the one for the $p$ quadrature measurement with a measurement outcome $\Im(y)$ is $W_{H_{\Im(y)}}(\alpha) = \frac{1}{\sqrt{2}\pi} \delta(\Im(\alpha)-\frac{\Im(y)}{\sqrt{2}})$. The factors of $\sqrt{2}$ are included such that we can rederive the ideal heterodyne detector POVM $\{E_y: y\in \mathbb{C}\}$ in the limit of unity detector efficiency and zero electronic noise. To discard modes $\gamma$ and $\omega$ that are not measured, we perform the integration over variables $\gamma$ and $\omega$. 

For any input state $\rho$ to the detector, one can in principle obtain the underlying probability density function $P(y)=\Tr(\rho G_y)$ for every measurement outcome $y \in \mathbb{C}$. As the correct POVM element $G_y$ needs to produce the observed probability density function $P(y)=\Tr(\rho G_y)$, this requirement in terms of Wigner functions becomes $P(y)= \pi \int d^2 \alpha W_{\rho} (\alpha)W_{G_y} (\alpha)$, where $W_{\rho}$ is the Wigner function of the input state $\rho$ and $W_{G_y}$ is the Wigner function of the operator $G_y$, by the overlap formula in Eq. (\ref{eq:overlap}). In Fig. \ref{fig:noisy_het_eq}, we know the mathematical description of measurements on the right, but the description of the state $W_{\text{out}}$ is unknown. On the other hand, we want to find the description of the measurement directly acting on the input state and the Wigner function description of the input state and those of ancillary modes on the left are either assumed to be given or known. To connect these known descriptions on the two sides of this diagram to find the desired Wigner function of the POVM element $G_y$ that acts on the input state directly, we start from the right-hand side of this diagram with an unknown four-mode state $W_{\text{out}}$ and the known measurements on these modes, perform inverse beam-splitter transformations from right to left of this diagram and finally obtain $W_{G_y}$ by integrating over variables other than $\alpha$. By starting with the multimode overlap formula for $P(y)$ on the right-hand side of the diagram and performing the process as described, we obtain
\begin{widetext}

\begin{aeq}\label{eq:wigner_firststep}
P(y) &= \pi^4 \int d^2 \alpha \int d^2 \beta \int d^2 \gamma \int d^2 \omega \; \frac{1}{\pi^2}W_{\text{out}}(\alpha,\beta,\gamma,\omega) W_{H_{\Re(y)}}(\alpha) W_{H_{\Im(y)}}(\beta) \\
&=\pi^2 \int d^2 \alpha  \; W_{\rho}(\alpha) \int d^2 \beta \; W_{\ket{0}}(\beta)  \int d^2 \gamma \; W_{\rho_{\text{th}}(\bar{n}_1)}(\gamma)  W_{H_{\Re(y)}}(\sqrt{\eta_1}\frac{\alpha+\beta}{\sqrt{2}}+\sqrt{1-\eta_1}\gamma) \\
&  \;\; \;\; \;\; \;\;\;\; \;\;  \;\; \;\; \;\; \;\;\;\; \;\; \;\; \;\; \;\; \;\;\;\; \;\;\times  \int d^2 \omega \; W_{\rho_{\text{th}}(\bar{n}_2)}(\omega)  W_{H_{\Im(y)}}(\sqrt{\eta_2}\frac{\alpha - \beta}{\sqrt{2}}+\sqrt{1-\eta_2}\omega).
\end{aeq}The next step is to substitute the Wigner function of the vacuum state in Eq. (\ref{eq:wigner_vacuum}) and that of the thermal state in Eq. (\ref{eq:wigner_thermal}) and then to perform the integrals over variables $\beta, \gamma$ and $\omega$. We first integrate over the variable $\omega$. The relevant integral that involves the variable $\omega$ is 
\begin{aeq}\label{eq:wigner_integration1}
 &\int d^2 \omega \; W_{\rho_{\text{th}}(\bar{n}_2)}(\omega)  W_{H_{\Im(y)}}(\sqrt{\eta_2}\frac{\alpha - \beta}{\sqrt{2}}+\sqrt{1-\eta_2}\omega)\\
 = & \frac{1}{\pi \sqrt{\pi}}\frac{1}{\sqrt{(1-\eta_2)(1+2\bar{n}_2)}} \exp(-\frac{\eta_2[\Im(\beta)+ \frac{1}{\sqrt{\eta_2}}\Im(y)-\Im(\alpha)]^2}{(1+2\bar{n}_2)(1-\eta_2)}).
\end{aeq}

Next, we perform the integral related to the variable $\gamma$. Since Eq. (\ref{eq:wigner_integration1}) does not involve the variable $\gamma$, we do not need to plug it back to solve the integral that involves the variable $\gamma$. This integration shown in Eq. (\ref{eq:wigner_integration2}) is actually similar to the integration that we just did in Eq. (\ref{eq:wigner_integration1}). 
\begin{aeq}\label{eq:wigner_integration2}
 &\int d^2 \gamma \; W_{\rho_{\text{th}}(\bar{n}_1)}(\gamma)  W_{H_{\Re(y)}}(\sqrt{\eta_1}\frac{\alpha+\beta}{\sqrt{2}}+\sqrt{1-\eta_1}\gamma)\\=& \frac{1}{\pi \sqrt{\pi}}\frac{1}{\sqrt{(1-\eta_1)(1+2\bar{n}_1)}}  \exp(-\frac{\eta_1 \big[\Re(\beta) - \frac{1}{\sqrt{\eta_1}}\Re(y)+\Re(\alpha) \big]^2}{(1+2\bar{n}_1)(1-\eta_1)}).
\end{aeq}Finally, we integrate over the variable $\beta$. We now need to substitute results of Eqs. (\ref{eq:wigner_integration1}) and (\ref{eq:wigner_integration2}) back to Eq. (\ref{eq:wigner_firststep}). The prefactor is simplified to be $\frac{1}{\pi^3}\frac{1}{\sqrt{(1-\eta_1)(1+2\bar{n}_1)(1-\eta_2)(1+2\bar{n}_2)}}$. Except this prefactor, we perform the following integral
\begin{aeq}\label{eq:wigner_integration3}
&\int d^2 \beta \; W_{\ket{0}}(\beta)  \exp(-\frac{\eta_1 \big[\Re(\beta) - \frac{1}{\sqrt{\eta_1}}\Re(y)+\Re(\alpha) \big]^2}{(1+2\bar{n}_1)(1-\eta_1)}) \exp(-\frac{\eta_2\big[(\Im(\beta)+ \frac{1}{\sqrt{\eta_2}}\Im(y)-\Im(\alpha)\big]^2}{(1+2\bar{n}_2)(1-\eta_2)}) \\
=&2\sqrt{\frac{(1+2\bar{n}_1)(1+2\bar{n}_2)(1-\eta_1)(1-\eta_2)}{(1+(1-\eta_1)+4\bar{n}_1(1-\eta_1))(1+(1-\eta_2)+4\bar{n}_2(1-\eta_2))}} \\
&\times \exp(\frac{-2 \eta_1 \big[\frac{1}{\sqrt{\eta_1}}\Re(y)-\Re(\alpha)\big]^2}{1+(1-\eta_1)+4 \bar{n}_1(1-\eta_1)}+\frac{-2 \eta_2 \big[\frac{1}{\sqrt{\eta_2}}\Im(y)-\Im(\alpha)\big]^2}{1+(1-\eta_2)+4\bar{n}_2(1-\eta_2)}). 
\end{aeq}

Finally, by putting the prefactor back and expressing the final expression in a format of Gaussian functions, we obtain the following result
\begin{aeq}\label{eq:wigner_Gy}
W_{G_y}(\alpha)
=&  \frac{1}{\sqrt{\eta_1\eta_2}\pi}\frac{2}{\pi}\frac{1}{\sqrt{1+\frac{2(1-\eta_1)(1+2\bar{n}_1)}{\eta_1}}}\frac{1}{\sqrt{1+\frac{2(1-\eta_2)(1+2\bar{n}_2)}{\eta_2}}} \\
& \times \exp(\frac{-2  [\frac{1}{\sqrt{\eta_1}}\Re(y)-\Re(\alpha)]^2}{1+\frac{2(1-\eta_1)(1+2\bar{n}_1)}{\eta_1}}+\frac{-2 [\frac{1}{\sqrt{\eta_2}}\Im(y)-\Im(\alpha)]^2}{1+\frac{2(1-\eta_2)(1+2\bar{n}_2)}{\eta_2}}).
\end{aeq}

By substituting in $\bar{n}_1 = \frac{\nu_{1}}{2(1-\eta_1)}$ and $\bar{n}_2 = \frac{\nu_{2}}{2(1-\eta_2)}$, we derive Eq. (\ref{eq:wigner_Gy_main}) after a straightforward simplification.

\subsection{POVM elements}
\subsubsection{General case}
As we derive the Wigner function of an arbitrary POVM element $G_y$ corresponding to the detector model in Fig. \ref{fig:noisy_het}, we next show that the POVM elements $G_y$'s are projections onto displaced squeezed thermal states up to a scaling factor. To see this, we make the following definitions:
\begin{aeq}\label{eq:parameters}
\lambda_j &:= \frac{(1-\eta_j)(1+2\bar{n}_j)}{\eta_j} = \frac{1-\eta_j+\nu_j}{\eta_j}  \; \; \text{   for } j = 1, 2,\\
\bar{n}_{het} &:= \frac{\sqrt{(1+2\lambda_1)(1+2\lambda_2)}-1}{2}, \\
\xi_{het} &:= \frac{1}{4}\ln(\frac{1+2\lambda_2}{1+2\lambda_1}), \\
\alpha_{het} &:= \frac{1}{\sqrt{\eta_1}}\Re(y)+\frac{i}{\sqrt{\eta_2}}\Im(y).
\end{aeq}
With these choices of parameters $\alpha_{het}, \xi_{het}$ and $\bar{n}_{het}$, Eq. (\ref{eq:wigner_Gy}) can be rewritten as
\begin{aeq}\label{eq:G_y_general}
W_{G_y}(\gamma)&=  \frac{1}{\sqrt{\eta_1\eta_2}\pi}\frac{2}{\pi}\frac{1}{1+2\bar{n}_{het}} 
 \exp{-2[\frac{e^{2\xi_{het}}\Re(\gamma-\alpha_{het})^2 + e^{-2\xi_{het}}\Im(\gamma-\alpha_{het})^2}{1+2\bar{n}_{het}}]}.
\end{aeq}
By comparing the Wigner function of $G_y$ in Eq. (\ref{eq:wigner_Gy}) and the Wigner function of a displaced squeezed thermal state in Eq. (\ref{eq:wigner_dsts}), we can identify $G_y = \frac{1}{\sqrt{\eta_1\eta_2}\pi} \rho_{\text{DSTS}}(\alpha_{het},\xi_{het},\bar{n}_{het})$ for the choices of parameters $\alpha_{het}, \xi_{het}$ and $\bar{n}_{het}$ in Eq. (\ref{eq:parameters}). Therefore, each $G_y$ is a scaled projection onto a displaced squeezed thermal state with the displacement $\alpha_{het}$, squeezing parameter $\xi_{het}$ and the mean photon number of the initial thermal state before squeezing and displacement $\bar{n}_{het}$.

\subsubsection{Simple case}
If $\eta_1 = \eta_2 = \eta_d$ and $\nu_{1} = \nu_{2} = \nu_{\text{el}}$, it is easy to verify that Eq. (\ref{eq:wigner_Gy}) reduces to Eq. (\ref{eq:wigner_Gy_main}). Then one can identify each POVM element $G_y$ is the projection onto a displaced thermal state with a prefactor $1/(\eta_d\pi)$ in Eq. (\ref{eq:noisy_het_povm}). An alternative view is to look at the parameters $\alpha_{het}, \xi_{het}$ and $\bar{n}_{het}$ in Eq. (\ref{eq:parameters}). In particular, since $\lambda_1 = \lambda_2$, the amount of squeezing $\xi_{het}$ becomes zero. Thus, by neglecting squeezing in the POVM elements of the general case, one can also conclude each POVM element is proportional to the projection onto a displaced thermal state. One can further verify that the displacement is $\alpha_{het} = \frac{y}{\sqrt{\eta_d}}$ and the mean photon number of the initial thermal state $\bar{n}_{het}$ becomes $\frac{1-\eta_d+\nu_{\text{el}}}{\eta_d}$.

\section{Photon-number basis representation of operators}\label{app:representation}
In this Appendix, we show how to represent region operators as well as observables needed for the optimization problem in Eq. (\ref{eq:optimization_reformulated}) in the photon-number basis. By the same discussion in Appendix B of Ref. \cite{Lin2019}, under the photon-number cutoff assumption, we need only to find $\Pi_{N} \hat{O} \Pi_{N}$ for an operator $\hat{O}$ with a cutoff photon number $N$. Thus, we are interested in finding the expression $\bra{m}\hat{O}\ket{n}$ for $0 \leq m, n \leq N$ for each relevant operator $\hat{O}$. When these operators are represented in this finite-dimensional Hilbert space spanned by $\{\ket{n}: 0 \leq n \leq N\}$, we can then proceed with numerical optimization to calculate the key rate.

In Sec. \ref{app_sec:simple_case}, our discussion is restricted to the simplified scenario that we use in the main text for presenting simulation results; that is, we set $\eta_1=\eta_2 =\eta_d$ and $\nu_1 =\nu_2 = \nu_{\text{el}}$. Under this scenario, we present formulae that can be efficiently evaluated in MATLAB. We then discuss the general case where the imperfections in two homodyne detectors are not necessarily the same in Sec. \ref{app_sec:general_case}. For the general case, we provide the matrix representation of the POVM elements $G_y$'s and leave the evaluation of region operators and observables for the optimization problem to be done numerically. 

\subsection{Simple case}\label{app_sec:simple_case}
In this simple case where two homodyne detectors in the heterodyne detection scheme have the same imperfection, the POVM element $G_y$ given in Eq. (\ref{eq:noisy_het_povm}) in the photon-number basis is expressed as \cite{Mollow1967}
\begin{aeq}\label{eq:G_y_off_diagonal}
\bra{m}G_y \ket{n} &= \frac{1}{\eta_d \pi} \exp[-\frac{\abs{y}^2}{\eta_d(1+\bar{n}_d)}] \frac{\bar{n}_d^m}{(1+\bar{n}_d)^{n+1}}(\frac{y^*}{\sqrt{\eta_d}})^{n-m} (\frac{m!}{n!})^{1/2} L_m^{(n-m)}\Big(-\frac{\abs{y}^2}{\eta_d \bar{n}_d(1+\bar{n}_d)}\Big),
\end{aeq}where we define $\bar{n}_d = \frac{1-\eta_d + \nu_{\text{el}}}{\eta_d}$ for ease of writing and $L_k^{(j)}(x)$ is the generalized Laguerre polynomial of degree $k$ with a parameter $j$ in the variable $x$. In particular, the diagonal entries are simplified to be
\begin{aeq}\label{eq:G_y_diagonal}
\bra{n}G_y \ket{n}  = \frac{1}{\eta_d \pi} \exp[-\frac{\abs{y}^2}{\eta_d(1+\bar{n}_d)}]\frac{\bar{n}_d^n}{(1+\bar{n}_d)^{n+1}}L_n(-\frac{\abs{y}^2}{\eta_d \bar{n}_d(1+\bar{n}_d)}),
\end{aeq}where $L_k(x) = L_k^{(0)}(x)$ is the Laguerre polynomial of degree $k$ in the variable $x$. For ease of writing later, we define $C_{m,n}= \frac{1}{ \pi \eta_d^{(n-m)/2+1} } (\frac{m!}{n!})^{1/2}  \frac{\bar{n}_d^m}{(1+\bar{n}_d)^{n+1}} $. 

\subsubsection{Region operators}
Our goal here is to write region operators $R_j = \int_{y \in \mathcal{A}_j} G_y \; d^2 y$ in the photon-number basis. For simplicity, we work out the expressions in the absence of postselection. To include the postselection, one may numerically integrate over the discarded region and subtract this result from the expression without postselection since this numerical integration is efficiently computable in MATLAB. We first consider off-diagonal elements (i.e. $m \neq n)$. In this case, we plug the expression of $\bra{m}G_y\ket{n}$ in Eq. (\ref{eq:G_y_off_diagonal}) into the definition of $R_j$ in Eq. (\ref{eq:region_operator_general}), write it in the polar coordinate with $y=re^{i\theta}$ and perform the integration over the phase $\theta$ to obtain the following expression:
\begin{aeq}
\bra{m}R_j\ket{n} &= \int_{0}^{\infty} dr \; r  \int_{(2j-1)\pi/4}^{(2j+1)\pi/4} d\theta \; \bra{m} G_{re^{i\theta}}\ket{n}\\
&=C_{m,n} \frac{i [e^{i(m-n)(2j-1)\pi/4}-e^{i(m-n)(2j+1)\pi/4}]}{m-n} \int_{0}^{\infty} dr \;  \exp[-\frac{r^2}{\eta_d(1+\bar{n}_d)}]  L_m^{(n-m)}\Big[-\frac{r^2}{\eta_d \bar{n}_d(1+\bar{n}_d)}\Big] r^{n-m+1}.
\end{aeq}Performing the integral over the variable $r$,  we obtain the result in terms of Taylor series expansion of a simple function:
\begin{aeq}
&\int_{0}^{\infty} dr \;  \exp[-\frac{r^2}{\eta_d(1+\bar{n}_d)}]  L_m^{(n-m)}\Big(-\frac{r^2}{\eta_d \bar{n}_d(1+\bar{n}_d)}\Big) r^{n-m+1}\\
 =& \; \; \frac{1}{2}[\eta_d (1+\bar{n}_d)]^{\frac{n-m}{2}+1}  \Gamma(\frac{n-m}{2}+1) f_m(\bar{n}_d,n-m,\frac{n-m}{2}),
\end{aeq}where $\Gamma$ is the gamma function and $f_m(a, \alpha, k)$ is defined as the Taylor series coefficients of the function below in the variable $t$ as
\begin{aeq}\label{eq:def_fn}
(1-t)^{-\alpha + k }(1-(1+\frac{1}{a})t)^{-(k+1)} = \sum_{n=0}^{\infty} f_n(a,\alpha , k) t^n.
\end{aeq}We note that the Taylor series coefficients here can be quickly found in MATLAB.

Now, we consider the diagonal entries of $R_j$ (i.e. $m=n$). 
By substituting $y=re^{i\theta}$ in Eq. (\ref{eq:G_y_diagonal}), we note that this expression does not depend on $\theta$. Thus, it is easy to see $\bra{n}R_0 \ket{n} =  \bra{n}R_1 \ket{n} = \bra{n}R_2 \ket{n} =\bra{n}R_3 \ket{n}.$ The integration over the phase $\theta$ gives a factor of $\frac{\pi}{2}$. We proceed the integration over variable $r$ and obtain
\begin{aeq}
\bra{n} R_j \ket{n} &= \frac{\pi}{2} \frac{1}{\eta_d \pi} \int_0^{\infty} dr\; r \exp[-\frac{r^2}{\eta(1+\bar{n}_d)}]\frac{\bar{n}_d^n}{(1+\bar{n}_d)^{n+1}}L_n\Big(-\frac{r^2}{\eta_d \bar{n}_d(1+\bar{n}_d)}\Big)=\frac{1}{4} \frac{\bar{n}_d^n}{(1+\bar{n}_d)^{n}}(1+\frac{1}{\bar{n}_d})^n=\frac{1}{4}.
\end{aeq}

To include postselection, the common integral in the case $m \neq n$ becomes 
\begin{aeq}
&\int_{\Delta_a}^{\infty} dr \;  \exp[-\frac{r^2}{\eta_d(1+\bar{n}_d)}]  L_m^{(n-m)}\Big(-\frac{r^2}{\eta_d \bar{n}_d(1+\bar{n}_d)}\Big) r^{n-m+1} \\=& \frac{1}{2}[\eta_d (1+\bar{n}_d)]^{\frac{n-m}{2}+1}  \Gamma(\frac{n-m}{2}+1) f_m(\bar{n}_d,n-m,\frac{n-m}{2})\\& -  \int_{0}^{\Delta_a} dr \;  \exp[-\frac{r^2}{\eta_d(1+\bar{n}_d)}]  L_m^{(n-m)}\Big(-\frac{r^2}{\eta_d \bar{n}_d(1+\bar{n}_d)}\Big) r^{n-m+1},
\end{aeq}where the second term is efficiently computable numerically. The case for $m=n$ follows similarly.
\subsubsection{First-moment observables}
We then proceed to evaluate the matrix elements of $\hat{F}_Q$ and $\hat{F}_P$. In the photon-number basis, the matrix elements are
\begin{aeq}
\bra{m}\hat{F}_Q \ket{n} &=  \int \frac{y+y^*}{\sqrt{2}} \bra{m}G_y\ket{n} \; d^2 y \\
&= \frac{C_{m,n}}{\sqrt{2}}  \int_{0}^{\infty} dr \; \exp[-\frac{r^2}{\eta_d(1+\bar{n}_d)}]  L_m^{(n-m)}\Big(-\frac{r^2}{\eta_d \bar{n}_d(1+\bar{n}_d)}\Big) r^{n-m+2} \int_{0}^{2\pi} d\theta \; e^{-i(n-m)\theta} (e^{i\theta}+e^{-i\theta}), \\
\bra{m}\hat{F}_P \ket{n} &=  \int \frac{i(y^*-y)}{\sqrt{2}} \bra{m}G_y\ket{n} \; d^2 y \\
&= \frac{i C_{m,n}}{\sqrt{2}}  \int_{0}^{\infty} dr \; \exp[-\frac{r^2}{\eta_d(1+\bar{n}_d)}]  L_m^{(n-m)}\Big(-\frac{r^2}{\eta_d \bar{n}_d(1+\bar{n}_d)}\Big) r^{n-m+2} \int_{0}^{2\pi} d\theta \; e^{-i(n-m)\theta} (e^{-i\theta}-e^{i\theta}). 
\end{aeq}As $\hat{F}_{Q}$ is a Hermitian operator, we can first find entries $\bra{m}\hat{F}_Q \ket{n} $ for $m \leq n$. Then for $m > n$, we simply set $\bra{m}\hat{F}_Q \ket{n} $ to be the complex conjugate of $\bra{n}\hat{F}_Q \ket{m}$. From the integration over $\theta$, the nonzero entries for $m \leq n$ are
\begin{aeq}
\bra{m}\hat{F}_Q \ket{m+1} &= \sqrt{2} \pi C_{m,m+1}  \int_{0}^{\infty} dr \; \exp[-\frac{r^2}{\eta_d(1+\bar{n}_d)}]  L_m^{(1)}\Big(-\frac{r^2}{\eta_d \bar{n}_d(1+\bar{n}_d)}\Big) r^{3} \\
&= \frac{\pi}{\sqrt{2}} C_{m,m+1} ((1+\bar{n}_d)\eta_d)^{2}  f_m(\bar{n}_d, 1, 1).\\
\end{aeq}By a similar procedure for $\hat{F}_P$, we have
\begin{aeq}
 \bra{m}\hat{F}_{P} \ket{m+1} &= - \sqrt{2} i \pi C_{m,m+1}  \int_{0}^{\infty} dr \; \exp[-\frac{r^2}{\eta_d(1+\bar{n}_d)}]  L_m^{(1)}\Big(-\frac{r^2}{\eta_d \bar{n}_d(1+\bar{n}_d)}\Big) r^{3} \\
&=-\frac{i \pi}{\sqrt{2}} C_{m,m+1} ((1+\bar{n}_d)\eta_d)^{2}  f_m(\bar{n}_d, 1, 1).
\end{aeq}

\subsubsection{Second-moment observables}
Next, we evaluate the matrix elements of $\hat{S}_Q$ and $\hat{S}_P$. In the photon-number basis, they are
\begin{aeq}
\bra{m} \hat{S}_{Q} \ket{n} &=  \int (\frac{y+y^*}{\sqrt{2}})^2 \bra{m}G_y\ket{n} d^2 y \\
&= \frac{C_{m,n}}{2} \int_{0}^{\infty}  dr \; \exp[-\frac{r^2}{\eta_d(1+\bar{n}_d)}]  L_m^{(n-m)}\Big(-\frac{r^2}{\eta_d\bar{n}_d(1+\bar{n}_d)}\Big) r^{n-m+3} \int_{0}^{2\pi} d\theta \; e^{-i(n-m)\theta} (e^{i\theta}+e^{-i\theta})^2, \\
\bra{m}\hat{S}_{P}\ket{n} &= \int (\frac{i(y^*-y)}{\sqrt{2}})^2 \bra{m}G_y\ket{n} d^2 y\\
&= -\frac{C_{m,n}}{2} \int_{0}^{\infty}  dr \; \exp[-\frac{r^2}{\eta_d(1+\bar{n}_d)}]  L_m^{(n-m)}\Big(-\frac{r^2}{\eta_d\bar{n}_d(1+\bar{n}_d)}\Big) r^{n-m+3} \int_{0}^{2\pi} d\theta \; e^{-i(n-m)\theta} (e^{-i\theta}-e^{i\theta})^2.
\end{aeq}Again, since $\hat{S}_Q$ and $\hat{S}_P$ are Hermitian operators, we only need to define the upper triangular part and then set the lower triangular part using the Hermitian property. The relevant integrals are simplified to be
\begin{aeq}
\bra{m} \hat{S}_{Q} \ket{m} &= 2 \pi  C_{m,m} \int_{0}^{\infty}  dr \; \exp[-\frac{r^2}{\eta_d(1+\bar{n}_d)}]  L_m \Big(-\frac{r^2}{\eta_d\bar{n}_d(1+\bar{n}_d)}\Big) r^{3}  \\
&= \pi  C_{m,m}(\eta_d(1+\bar{n}_d))^2 f_m(\bar{n}_d, 0, 1),\\
\bra{m} \hat{S}_{Q} \ket{m+2} &=  \pi C_{m,m+2} \int_{0}^{\infty}  dr \; \exp[-\frac{r^2}{\eta_d(1+\bar{n}_d)}]  L_m^{(2)}\Big(-\frac{r^2}{\eta_d\bar{n}_d(1+\bar{n}_d)}\Big) r^{5} \\
&=\pi C_{m,m+2} (\eta_d(1+\bar{n}_d))^3 f_m(\bar{n}_d, 2, 2).
\end{aeq}

 For $\hat{S}_{P}$, we have
\begin{aeq}
\bra{m} \hat{S}_{P} \ket{m} &= 2 \pi  C_{m,m} \int_{0}^{\infty}  dr \; \exp[-\frac{r^2}{\eta_d(1+\bar{n}_d)}]  L_m \Big(-\frac{r^2}{\eta_d\bar{n}_d(1+\bar{n}_d)}\Big) r^{3}  \\
&= \pi  C_{m,m}(\eta_d(1+\bar{n}_d))^2 f_m(\bar{n}_d, 0, 1),\\
\bra{m} \hat{S}_{P} \ket{m+2} &= - \pi C_{m,m+2} \int_{0}^{\infty}  dr \; \exp[-\frac{r^2}{\eta_d(1+\bar{n}_d)}]  L_m^{(2)}\Big(-\frac{r^2}{\eta_d\bar{n}_d(1+\bar{n}_d)}\Big) r^{5} \\
&=-\pi C_{m,m+2} (\eta_d(1+\bar{n}_d))^3 f_m(\bar{n}_d, 2, 2).
\end{aeq}

\subsection{General case}\label{app_sec:general_case}
We consider the general case where two homodyne detectors may have different imperfections. In this case, each POVM element $G_y$ is given in Eq. (\ref{eq:G_y_general}). Given the POVM $G_y =  \frac{1}{\sqrt{\eta_1\eta_2}\pi} \rho_{\text{DSTS}}(\alpha_{het},\xi_{het},\bar{n}_{het}),$ its matrix elements are given by Eq. (5.2) of Ref. \cite{Marian1993} with the prefactor $\frac{1}{\sqrt{\eta_1\eta_2}\pi}$ as
\begin{aeq}\label{eq:general_Gy_entry}
\bra{m}G_y\ket{n} &= \frac{1}{\sqrt{\eta_1\eta_2}} \frac{Q(0)}{\sqrt{m!n!}}\sum_{k=0}^{\min(m,n)} k! \binom{m}{k} \binom{n}{k} \tilde{A}^k \Big(\frac{\tilde{B}}{2}\Big)^{\frac{m-k}{2}} \Big(\frac{\tilde{B}^*}{2}\Big)^{\frac{n-k}{2}}  H_{m-k}((2\tilde{B})^{-\frac{1}{2}}\tilde{C}) H_{n-k}((2\tilde{B}^*)^{-\frac{1}{2}}\tilde{C}^{*}),
\end{aeq}where $H_{\ell}$ is the Hermite polynomial of order $\ell$. With simple substitutions, one may verify that these parameters $\tilde{A}$, $\tilde{B}$, $\tilde{C}$ and $Q(0)$ are defined in terms of $\lambda_1, \lambda_2, \alpha_{het}$ in Eq. (\ref{eq:parameters}) as 
\begin{aeq}
\tilde{A} &:= 1- \frac{\lambda_1+\lambda_2+2}{2(\lambda_1+1)(\lambda_2+1)},\\
\tilde{B} &:= \frac{-\abs{\lambda_1-\lambda_2}}{2(\lambda_1+1)(\lambda_2+1)},\\
\tilde{C} &:= \frac{\Re(\alpha_{het})}{\max(\lambda_1, \lambda_2) +1} + i \frac{\Im(\alpha_{het})}{\min(\lambda_1, \lambda_2)+1},\\
Q(0) &:= \frac{1}{\pi} \frac{1}{\sqrt{(\lambda_1+1)(\lambda_2+1)}}\exp[-\frac{\Re(\alpha_{het})^2}{\max(\lambda_1, \lambda_2)+1}-\frac{\Im(\alpha_{het})^2}{\min(\lambda_1, \lambda_2)+1}].
\end{aeq}As indicated in Eqs. (5.3) and (5.4) of Ref. \cite{Marian1993}, the choice of square roots of $\tilde{B}$ is as $\tilde{B}^{1/2} = ie^{i(\varphi/2)} \abs{\tilde{B}}^{1/2}$ and $(\tilde{B}^*)^{1/2} =(\tilde{B}^{1/2})^*$, where $\varphi = 0$ if $\lambda_1 \leq \lambda_2$ and $\varphi = \pi$ if $\lambda_1 > \lambda_2$. We note that $\bar{n}_{het}$ and $\xi_{het}$ are defined in terms of $\lambda_1$ and $\lambda_2$ and one may rewrite these parameters in terms of $\bar{n}_{het}, \xi_{het}$ and $\alpha_{het}$ to make the matrix elements more explicitly depend on the parameters of the displaced squeezed thermal states.

From the expression of $\bra{m}G_y\ket{n}$ in Eq. (\ref{eq:general_Gy_entry}), one can apply the definition of region operators $R_j$'s  in Eq. (\ref{eq:region_operator_general}) to find  $\bra{m}R_j\ket{n}$ by numerical integration. Similarly, from the definitions of first- and second-moment observables in Eq. (\ref{eq:observables}) in terms of POVM elements $G_y$'s, one can numerically obtain a representation of these operators in the photon-number basis. 
\end{widetext}
\bibliographystyle{apsrev4-2}
\bibliography{DMCVTN}

\end{document}